\documentclass[11pt,a4paper]{article}
\pdfoutput=1 
\usepackage{amsmath,amssymb,amscd,amsxtra}
\usepackage[mathscr]{eucal}
\usepackage{jheppub}
\usepackage{bigints}
\usepackage{color}
\usepackage{latexsym}
\usepackage{ytableau}
\usepackage{mathtools}   
\usepackage{amsfonts}
\usepackage{ifthen}
\usepackage[]{hyperref} 
\usepackage{bbm}
\usepackage{amsthm}
\usepackage{tikz}
\newtheorem{conjecture}{Conjecture}
\usetikzlibrary{arrows}

\makeatletter
\pgfdeclareshape{datastore}{
  \inheritsavedanchors[from=rectangle]
  \inheritanchorborder[from=rectangle]
  \inheritanchor[from=rectangle]{center}
  \inheritanchor[from=rectangle]{base}
  \inheritanchor[from=rectangle]{north}
  \inheritanchor[from=rectangle]{north east}
  \inheritanchor[from=rectangle]{east}
  \inheritanchor[from=rectangle]{south east}
  \inheritanchor[from=rectangle]{south}
  \inheritanchor[from=rectangle]{south west}
  \inheritanchor[from=rectangle]{west}
  \inheritanchor[from=rectangle]{north west}
  \backgroundpath{
    \southwest \pgf@xa=\pgf@x \pgf@ya=\pgf@y
    \northeast \pgf@xb=\pgf@x \pgf@yb=\pgf@y
    \pgfpathmoveto{\pgfpoint{\pgf@xa}{\pgf@ya}}
    \pgfpathlineto{\pgfpoint{\pgf@xb}{\pgf@ya}}
    \pgfpathmoveto{\pgfpoint{\pgf@xa}{\pgf@yb}}
    \pgfpathlineto{\pgfpoint{\pgf@xb}{\pgf@yb}}
 }
}
\makeatother

\begin{document}

\title{The Euler anomaly and scale factors in Liouville/Toda CFTs}
\author{Aswin Balasubramanian}
\affiliation{Theory Group\\ Department of Physics \\ University of Texas at Austin\\2515 Speedway Stop C1608\\
Austin, TX 78712-1197}
\emailAdd{aswin@utexas.edu}
\abstract{ The role played by the Euler anomaly in the dictionary relating sphere partition functions of four dimensional theories of class $\mathcal{S}$ and two dimensional non rational CFTs is clarified. On the two dimensional side, this involves a careful treatment of scale factors in Liouville/Toda correlators. Using ideas from tinkertoy constructions for Gaiotto duality, a framework is proposed for evaluating these scale factors. The representation theory of Weyl groups plays a critical role  in this framework.}
\keywords{supersymmetric field theories, representation theory, conformal field theory.}

\begin{titlepage}

\begin{flushright}
UTTG-28-13\\
\end{flushright}
\vskip 1cm

\maketitle
\end{titlepage}

\section{Introduction}

In recent investigations of the dynamics of a class of $\mathcal{N} =2$ superconformal field theories in four dimensions (often called theories of class $\mathcal{S}$ to underline their six dimensional origin), it has become increasingly clear that various observables of this class of theories admit an efficient description using the language of two dimensional physics. A particular example of such an observable is the partition function of the four dimensional theory defined on a sphere ($Z_{\mathbb{S}^4}$).  Following Pestun's evaluation of the partition function for a subset of class $\mathcal{S}$ theories theories via localization \cite{Pestun:2007rz} and the construction of these theories using the $(0,2)$ six dimensional theory SCFT $\mathscr{X}[\mathfrak{g}]$ \footnote{There is a unique six dimensional theory for every choice of $\mathfrak{g} \in {A,D,E}$. } \cite{Gaiotto:2009we, Gaiotto:2009hg}, AGT noticed the remarkable fact that the partition functions in type $\mathfrak{g}=A_1$ coincide with certain correlators in a particular Liouville conformal field theory \cite{Alday:2009aq}. They further conjectured (see also \cite{Wyllard:2009hg} in this regard) that an analogous relationship exists for partition functions of various higher rank theories and corresponding Toda correlators. Many checks of this proposal are available in specific corners of the moduli space where the four dimensional theories admit a Lagrangian description as a weakly coupled gauge theory along with conventional matter multiplets. At other corners  of the moduli space (which happen to be the vast majority), one runs into the following predicament. On the four dimensional side, the localization techniques do not extend as there is no known Lagrangian description. On the two dimensional side, a complete analytical understanding of the corresponding Toda correlators is missing. One of the initial motivations for this work was to partly alleviate this situation by pointing out that the AGT dictionary can very easily be expanded to include an observable that is much better understood, namely the Euler anomaly of the four dimensional SCFT.  Borrowing ideas from the tinkertoy constructions, I propose a framework for calculating this dependence. This framework is of independent interest and can potentially shine light on certain aspects of the tinkertoy constructions.  This paper is confined to theories of type $A_n$. Investigations in more general cases will be reported elsewhere.

Here is a short outline of the paper. In Section \ref{partitioneuler}, the encoding of the Euler anomaly in the sphere partition function is described. In Section \ref{liouvilleeuler} the corresponding scale factor on the Liouville side for the $A_1$ theories is identified. Sections \ref{Antheories1}, \ref{Antheories2} generalize these arguments to the case of $A_n$ Toda theories. Here, it is argued that the tinkertoys of \cite{Chacaltana:2010ks} have natural analogs in the world of Toda CFTs. Using this dictionary of tinkertoys, one can calculate the scale factors for many Toda correlators. When the Toda correlator corresponds to a free fixture, this gives a prediction regarding the analytical structure of the corresponding Toda three point function. An important element of this calculation of scale factors is a realization of the relevant Toda primaries starting from primaries in a WZW model. This map between the primaries is further helpful in elucidating the relationship to the theory of nilpotent orbits in semisimple Lie algebras and the closely related theory of Weyl group representations. A degree of familiarity with the relevant class $\mathcal{S}$ constructions \cite{Gaiotto:2009hg,Gaiotto:2009we,Chacaltana:2010ks,Gaiotto:2011xs,Chacaltana:2012zy,Nanopoulos:2009uw} will greatly help in the reading of the paper. 

The class of theories studied here have recently attracted much attention from various physical and mathematical perspectives. While connections to some of these are discussed in the final section, it would be foreboding to try and comment on all of them. Instead, the interested reader is referred to some reviews-in-preparation \cite{mooreKleinLecs,TachikawaLecs,TachikawaReview} for such overviews.

\section{Partition function on $\mathbb{S}^4$ and the Euler anomaly}
\label{partitioneuler}

It is expected that the logarithm of the sphere partition function has a divergent piece that is proportional to the Euler anomaly $a$ \cite{Cardy:1988cwa}. This is an important observable for any CFT since it is a measure of the massless degrees of freedom in the CFT. In \cite{Cardy:1988cwa}, it was also conjectured that such a measure exists at all points along a renormalization group flow and that its value strictly decreases as more degrees of freedom are integrated out. A version of this conjecture has recently been proved in \cite{Komargodski:2011vj}.  The goal here is to focus on the class $\mathcal{S}$ SCFTs and make the dependence on the Euler anomaly manifest in their sphere partition functions. We will begin by considering the case of conformal class $\mathcal{S}$ theories with Lagrangian descriptions.  A definition of these theories on the round four sphere and a localization scheme to evaluate the partition function of the theory so defined\footnote{See also \cite{Festuccia:2011ws} and \cite{Dumitrescu:2012ha} on the question of defining such theories on curved manifolds.} was described by Pestun \cite{Pestun:2007rz}. This construction was recently extended to the case of the more general case of an ellipsoid $\mathbb{S}^4_b$ \cite{Hama:2012bg}. In much of the literature on the AGT conjecture, the dependence of the partition function on the Euler anomaly is not made explicit\footnote{For considerations of similar issues in three dimensions, see \cite{Jafferis:2010un}.}. In the original work of \cite{Pestun:2007rz}, this was not necessary as the corresponding factors in the partition function cancel in the calculation of expectation values of BPS Wilson and 't-Hooft loop operators\footnote{I thank V.Pestun for a discussion.}. For the purposes of this work, it would be important to make this dependence explicit. This paper will be restricted to analyzing the case of a round sphere.

While this paper will focus solely on the physical $\mathcal{N}=2$ theories, it is interesting to note that the dependence made explicit here has a cousin in the world of topological QFTs obtained from twisting the Lagrangian $\mathcal{N}=2$ theories. In the evaluation of their partition functions on a general four manifold (with non-zero Euler characteristic $\chi$ and signature $\sigma$), the measure in the path integral has an explicit dependence on the anomaly parameters $a, c$ \cite{Shapere:2008zf}.

\subsection{Localization on the four sphere}

For a general superconformal $\mathcal{N}=2$ theory with matter in representation $W$ of the gauge group $G$ taken on a sphere $\mathbb{S}^4$ of radius $R_0$, the one loop functional determinant around the locus of classical solutions on which the theory localizes was evaluated in \cite{Pestun:2007rz}. It takes the following form,
\begin{equation*}
Z_{1-loop}^{W} = \frac{\prod_{\alpha \in \text{weights(Ad)}}\prod_{n=1}^{\infty}((\alpha.a_E)^2 + \mu^2  n^2)^n}{\prod_{ w\in \text{weights(W)}} \prod_{n=1}^{\infty}((w.a_E)^2 + \mu^2  n^2)^n}.
\label{oneloop}
\end{equation*}
The hypermultiplet masses have been set to zero and $\mu=R_0^{-1}$. Let us focus our attention on a prototypical infinite product that occurs in these determinants and go through with the steps of regularizing it. We choose the one in the numerator of the example just studied and rewrite it as 
\begin{equation*}
\prod_{n=1}^{\infty}((\alpha.a_E)^2 + \mu^2 n^2)^n = \prod_{n=1}^{\infty} (i(\alpha.a_E) + \mu  n)^n(-i(\alpha.a_E) + \mu  n)^n.
\end{equation*}
Each factor can further be rewritten as 
\begin{equation}
\prod_{n=1}^{\infty} (i(\alpha.a_E) + \mu  n)^n = \frac{\prod_{n,m \in N^2} (i(\alpha.a_E) + \mu  m + \mu n)}{\prod_{n\in N}(i(\alpha.a_E) + n \mu )},
\label{toregulate}
\end{equation}
where $N^2$ is the set of all $(m,n)$ such that $m,n \in \mathbb{N} = {0,1,2, \ldots} $. The form of the infinite product in the numerator is very suggestive of a regularizing scheme using the Barnes double zeta function $\zeta_2^B$. For the denominator, the Hurwitz zeta function seems like the appropriate choice. Let us recall the sum representation for $\zeta_2^B$,
\begin{equation*}
\zeta_2^B(s,x;a,b) = \sum_{m=0,n=0}^{m=\infty,n=\infty} (x+ a m + b n)^{-s}.
\end{equation*}
$\zeta_2^B (s,x)$ can be analytically continued to a meromorphic function which has poles when $x= -n_1 a - n_2 b$. We can use $\zeta_2^B$ to regulate infinite products using the following (formal) identity
\begin{equation*}
   \prod_{n,m \in N_0 } (x + ma + nb)   = e^{-\zeta_2^{B'} (0,x;a,b)} .
\end{equation*}
Before the products in this problem are regularized, it is helpful to note that under a scaling transformation that takes $(x,a,b )\rightarrow (kx,ka,kb)$, the new regularized product is related to old product in the following way (the additional steps are reviewed in Appendix \ref{appendixscaling})
\begin{equation*}
 \prod_{n,m=0}^{\infty} (k(x + ma + nb)) = k^{\zeta_2^{B} (0,x;a,b)} e^{-\zeta_2^{B'} (0,x;a,b)}.
\end{equation*}
Similar equations hold for the Hurwitz zeta function.
Now, using $x= i(\alpha.a_E), k=\mu, a= 1, b= 1$, (\ref{toregulate}) is regularized to
\begin{equation*}
\frac{\prod_{n,m \in N^2} (i(\alpha.a_E) + \mu  m + \mu  n)}{\prod_{n\in N}(i(\alpha.a_E) + n \mu )} = \mu^{\zeta_2^{B} (0,i(\alpha.a_E);1,1)-\zeta^H(0,i(\alpha.a_E))} e^{-\zeta_2^{B'} (0,i(\alpha.a_E);1,1)+\zeta^{H'}(0,i(\alpha.a_E))}.
\end{equation*}
Further noting that
\begin{eqnarray*}
\zeta_2^{B} (0,x;1,1) &=& \frac{5}{12} - x + \frac{x^2}{2},   \\ \zeta^H(0,x) &=& \frac{1}{2} - x,
\end{eqnarray*}
and 
\begin{equation*}
e^{-\zeta_2^{B'} (0,x;1,1)+\zeta^{H'}(0,x;1,1)} = G (1+x),
\end{equation*}
where $G(z)$ is the Barnes G function, \footnote{For a summary of properties of the Barnes function and other special functions that appear in the paper, see Appendix \ref{appendixspecial}.} the regularized product becomes
\begin{equation*}
\mu^{-\frac{1}{12} + \frac{ \alpha.a_E^2}{2}} G(1+  \frac{i \alpha.a_E}{\mu}).
\end{equation*}
Thus the total contribution from each root in \ref{oneloop} is
\begin{equation*}
\mu^{-\frac{1}{6} + (\alpha.a_E^2)^2} G(1+  \frac{i \alpha.a_E}{\mu})G(1 -  \frac{i \alpha.a_E}{\mu}).
\end{equation*}
Regulating each piece in a similar way and defining $H(z) = G(1+z)G(1-z)$, 
\begin{equation*}
Z_{1-loop}^{W} = \frac{\prod_{\alpha \in \text{weights(Ad)}}\mu^{-1/6}H(\frac{i \alpha.a_E}{\mu})}{\prod_{ w\in \text{weights(W)}} \mu^{-1/6} H(\frac{i w.a_E}{\mu})}.
\end{equation*}
In the above step, the expression has been simplified using the condition for vanishing beta function
\begin{equation*}
\sum_{\alpha \in \text{weights(Ad)}} (\alpha.a_E)^2 = \sum_{w\in \text{weights(W)} }(w.a_E)^2.
\end{equation*} 
Let us specialize to the case of $G=SU(N)$ and $N_f=2N$. This gives,
\begin{equation}
Z_{1-loop,SU(N)}^{N_f=2N} = \mu^{\frac{1}{6}(N^2 +1)}\frac{\prod_{\alpha \in \text{weights(Ad)}}H(\frac{i \alpha.a_E}{\mu})}{\prod_{ w\in \text{weights(W)}}  H(\frac{i w.a_E}{\mu})}.
\label{oneloopscaled}
\end{equation}
The $\mu$ dependent factor in front of the product of $H$ functions in (\ref{oneloopscaled}) will play an important role in the identification of the Euler anomaly in the next section.
\subsection{The Euler anomaly}
All the necessary tools required bring out the dependence of the sphere partition function on the Euler anomaly are now assembled. From \cite{Pestun:2007rz}, the general form of the partition function (including non-perturbative contributions) is 
\begin{equation*}
Z_{\mathbb{S}^4} = \int_{\mathbf{a} \in \mathfrak{g}} d \mathbf{a} e^{-S_{cl}(\mathbf{a}, \mu )} Z_{1-loop} (\mathbf{a}, \mu) | Z_{inst} (\mathbf{a}, \mu) |^2,
\end{equation*}
where $S_{cl} = \frac{8 \pi (\mathbf{a},\mathbf{a})}{g^2 \mu^2}$ and $Z_{1-loop}$ is given by (\ref{oneloopscaled}). $Z_{inst}$ is the Nekrasov partition function defined on a $\Omega_{\epsilon_1, \epsilon_2}-$ background with $\epsilon_1 = \epsilon_2 = \mu$. 
This can be reduced to an integral over the Cartan subalgebra $\mathfrak{h}\subset\mathfrak{g}$ 
\begin{equation}
Z_{\mathbb{S}^4} = \int_{\mathbf{a} \in \mathfrak{h}}  d \mathbf{a} \mathbf{V}(\mathbf{a}) e^{-S_{cl}(\mathbf{a}, \mu )} Z_{1-loop} (\mathbf{a}, \mu) | Z_{inst} (\mathbf{a}, \mu) |^2,
\end{equation}
where  $\mathbf{V}(\mathbf{a})$ is the Vandermonde determinant.
 It is now convenient to change variable from $\mathbf{a}$ to  $\mathbf{\tilde{a}=\mathbf{a}/\mu}$. Note here that the form of $S_{cl}$ and $Z_{inst}$ are such that they are independent of $\mu$ when expressed in terms of $\tilde{\mathbf{a}}$. So, the integral in the new variables is 
\begin{equation}
 Z_{\mathbb{S}^4} = \mu^{(N^2-1) + \frac{1}{6} (N^2+1)} \int_{\mathbf{\tilde{a}} \in \mathfrak{h}} d \mathbf{\tilde{a}} \mathbf{V}(\mathbf{\tilde{a}}) e^{-S_{cl}(\mathbf{\tilde{a}} )} Z_{1-loop} (\mathbf{\tilde{a}}) | Z_{inst} (\mathbf{\tilde{a}}) |^2.
\label{anomaly}
\end{equation}
The exponent of $\mu$ in the above expression can be identified as $4 a$ where $a$ is the Euler anomaly of the theory. This factor should be proportional to $\chi a$ where $\chi$ is the Euler characteristic of the curved manifold on which the theory is defined. To fix conventions concretely, one can follow \cite{Duff:1977ay} and set
\begin{equation}
 Z^{-1}\mu \frac{\partial Z}{\partial \mu} = - \int dx^4 \langle T^j_j \rangle = 2 \chi a.
 \label{duff}
\end{equation}
In a theory with $N_S$ real scalars, $N_F$ Dirac fermions and $N_V$ vector fields, $a$ (as normalized above) is given by 
\begin{equation}
a = \frac{1}{360} (N_S + 11 N_F + 62 N_V).
\end{equation}
Recall that a $\mathcal{N}=2$ vector multiplet is the equivalent of a vector field, two real scalars and a single Dirac fermion and that a $\mathcal{N}=2$ hypermultiplet is the equivalent of four real scalars and one Dirac fermion. So, for a $\mathcal{N}=2$ theory with $n_v$ vector multiplets and $n_h$ hyper multiplets, 
\begin{equation}
4a = n_v + \frac{n_h - n_v}{6}.
\end{equation}
From \ref{anomaly}, calculate
\begin{equation}
Z^{-1}\mu \frac{\partial Z}{\partial \mu} = (N^2-1) + \frac{N^2+1}{6}.
\label{eulercharge}
\end{equation}
and note that the result equals $4a$ for the theory. Noting that $\chi(\mathbb{S}^4)=2$, this indeed matches with \ref{duff}.
For Lagrangian theories (like the ones considered so far), parameterizing $a$ by $n_v$, $n_h$ is the most obvious choice for these correspond to the number of vector multiplets and the number of hypermultiplets. Often, this is used for arbitrary theories with the understanding that it is just a convenient parameterization of the trace anomalies. It is then appropriate to call $n_h$ and $n_v$ the effective number of hypermultiplets and vector multiplets. The formula for the other trace anomaly $c$ is given by
\begin{equation}
c= \frac{n_v}{4} + \frac{n_h-n_v}{12}.
\end{equation}
For a general class $\mathcal{S}$ theory obtained by taking  theory $\mathscr{X}[\mathfrak{g}]$ on $C_{g,n}$,  the quantities $n_v$ and $n_h - n_v$ are related to the dimensions of vacuum moduli spaces in a simple fashion. Let $d_k$ denote the graded Coulomb branch dimension, that is the number of Coulomb branch operators of degree $k$. $n_v$ is given by
\begin{equation}
n_v = \sum_k (2k-1) d_k.
\end{equation}
$(n_h -n_v)$ on the other hand is  equal to the quaternionic Higgs branch dimension when there is such a branch. For theories without a true Higgs branch, one can still associate a maximally Higgsed branch whose quaternionic dimension is $n_h - n_v$ upto some abelian vector multiplets\cite{Gaiotto:2011xs},
\begin{equation}
\text{dim}_\mathbb{Q} (\mathcal{H}) = n_h -n_v + g \textbf{rank}(\mathfrak{g}).
\end{equation}
The total $n_h$ and $n_v$ for any theory is computed as in \cite{Chacaltana:2012zy},
\begin{eqnarray}
n_h = \sum_i n_h ^i + n_h ^{global}, \\
n_v = \sum_i n_v^i + n_v^{global},
\label{NhandNv}
\end{eqnarray}
where the global contributions \footnote{The central charge of the Toda CFT of type $\mathfrak{g}$ also has a similar presentation owing to the fact that it too can be obtained from the anomaly polynomial in six dimensions\cite{Bonelli:2009zp,Alday:2009qq}.} are given by \cite{Benini:2009mz,Alday:2009qq}
\begin{eqnarray}
n_h ^{global} &=& \frac{4}{3}(g-1) \hat{h} (\text{dim} G ), \nonumber \\
n_v^{global} &=& (g-1) (\frac{4}{3} \hat{h} \text{dim} G + \text{rank} G ),
\label{NhNvformulae}
\end{eqnarray}
where $\hat{h}$ is the dual Coxeter number and $n_h^i,n_v^i$ are the local contributions from a codimension two defect. In the rest of the paper, the goal will be to understand how the Euler anomaly (\ref{eulercharge}) is encoded in the Liouville/Toda correlators assigned to a general class $\mathcal{S}$ theory of type $\mathfrak{g} = A_n$.
\section{Scale factors in Liouville correlators} 
In this section, the prefactor that encodes the Euler central charge is shown to have a natural role in Liouville theory. It will be identified with the scale factor for the stripped correlator. A plausible path integral argument for how this scale factor arises is provided for the simplest case of a three point function and will be used to get some intuition for the appearance of such a factor. For higher point functions, such a luxury does not exist and one would have to resort to calculating them directly from the scaling behaviour of the $\Upsilon$ functions that occur in the DOZZ formula.
\label{liouvilleeuler}

Recall that Liouville field theory on a Riemann surface $C$ is defined by the following action (written with an unconventional normalization, $\phi = \hat{\phi}/6$ where $\hat{\phi}$ is the Liouville field in the usual normalization),
\begin{equation}
S_L = \frac{1}{72 \pi} \int \sqrt{\hat{g}} d^2 z \bigg ( \frac{1}{2}\hat{g}^{ab} \partial_a \phi \partial_b \phi + 3 Q \hat{R} \phi + 2 \pi \Lambda e^{2b\phi}  \bigg),
\end{equation}
where $z$ is a complex co-ordinate on the $C$. This theory is conformal upto a $c-$ number anomaly. While the observables of the theory depend only on the conformal class of the metric $g$ on $C$, it is often convenient to perform calculations by choosing a particular reference metric $\hat{g}$ in the same conformal class as $g$. The action above is written in terms of this reference metric. The physical metric is given by $g_{ab} = e^{\frac{2 \hat{\phi}}{Q}} \hat{g}_{ab}$.  The stress energy tensor for this theory is a shifted version of that for a free theory :
\begin{equation}
T(z) = - (\partial \hat{\phi})^2 + Q\partial^2 \hat{\phi}
\end{equation}
and the central charge is given by
\begin{equation}
c= 1 + 6 Q^2.
\end{equation}
Let us now formulate this theory on the Euclidean two sphere. Here, $g$ is taken to be the usual round metric and $\hat{g}$ as a flat metric. Calculations with the reference metric are to be done with the understanding that there is an operator insertion at infinity that encodes the curvature of the physical metric. A way to demand this is through a boundary condition for the field $\phi$
\begin{equation}
\phi = - 2Q \log (R/R_0) + \mathcal{O} (1),
\label{boundarycondition}
\end{equation}
where $R$ (=$\sqrt{z \bar{z}}$) is the distance measured in the flat reference metric. The parameter $R_0$ is introduced here for purely dimensional reasons. Its role in the overall scheme of things will become more transparent as we proceed. Now, a way to restrict to an integration over only fields that obey (\ref{boundarycondition}) is to write the Liouville action on a disc of radius $R$ along with boundary term that implements the curvature boundary condition and a field independent term that keeps the action finite in the $R \rightarrow \infty$ limit\footnote{Henceforth, such a limit will be assumed whenever Liouville/Toda actions on the disc are considered.}.
\begin{equation}
S_{L,\text{disc}} =  \frac{1}{72 \pi} \int_D \sqrt{\hat{g}} d^2 z \bigg ( \frac{1}{2} \hat{g}^{ab} \partial_a \phi \partial_b \phi   + 2 \pi \Lambda e^{2b\phi}  \bigg) + \frac{Q}{12\pi}\int_{\partial D} \phi d \theta + \frac{1}{3} Q^2 \log (R/R_0).
\label{action}
\end{equation}
The above action is invariant under a conformal transformation of the metric combined with a corresponding shift in the Liouville field,
\begin{eqnarray*}
z' &=& w(z), \\
\phi'(z) &=& \phi(z) - \frac{Q}{2} \log \bigg( \frac{\partial w}{\partial z} \bigg)^2.
\end{eqnarray*}
Note that last term plays an important role in ensuring invariance under this transformation and further, it also guarantees that the action is finite \cite{Harlow:2011ny,Zamolodchikov:1995aa}. 

According to the AGT correspondence, the partition function of a $A_1$ class $\mathcal{S}$ theory on the round sphere is identified with a corresponding $n-$point correlator in the $c=25$ Liouville CFT (upto some factors). Recall that these theories are obtained by compactifying theory $\mathscr{X}[\mathfrak{g}]$ on a Riemann surface $C_{g,n}$ of genus $g$ in the presence of $n$ codimension two defects whose locations on $C$ are given by $n$ punctures. The AGT correspondence assigns to this theory a Liouville correlator $\langle \mathcal{O}_1 \ldots \mathcal{O}_n \rangle$ where $\mathcal{O}_i= e^{2 \alpha_i \phi}$. The Liouville momenta are related to the mass deformation parameters of the 4d theory as $\alpha_i = Q/2 + i m_i$. One of the simplest examples of this 4d-2d dictionary is illustrated by the case of a sphere with three punctures. This corresponds to a theory of four free hypermultiplets. On the Liouville side, the correlator is known to take the following form,
\begin{equation*}
 V[\mathfrak{sl_2}]_{0,([1^2],[1^2],[1^2])} = C(\alpha_1,\alpha_2,\alpha_3) |z_{12}|^{-2(\Delta_1 + \Delta_2 -\Delta_3)} |z_{13}|^{-2(\Delta_1 + \Delta_3 - \Delta_2)} |z_{23}|^{-2(\Delta_2 + \Delta_3 -\Delta_1)}.
\end{equation*}
where $C(\alpha_1,\alpha_2,\alpha_3)$ is given by,
\begin{eqnarray*}
 C(\alpha_1,\alpha_2,\alpha_3) &=&  \bigg[\pi \Lambda  \gamma(b^2) b^{2-2b^2} \bigg]^{(Q-\sum_i \alpha_i)/b}  \times \\ &&\frac{\Upsilon(b) \Upsilon(2 \alpha_1)\Upsilon(2\alpha_2)\Upsilon(2\alpha_3)}{\Upsilon(\alpha_1 + \alpha_2 + \alpha_3 - Q)\Upsilon(\alpha_1 + \alpha_2 - \alpha_3)\Upsilon(\alpha_2 + \alpha_3 - \alpha_1)\Upsilon(\alpha_3 + \alpha_1 - \alpha_2)}.
\end{eqnarray*}

\begin{figure}
\begin{center}
\includegraphics{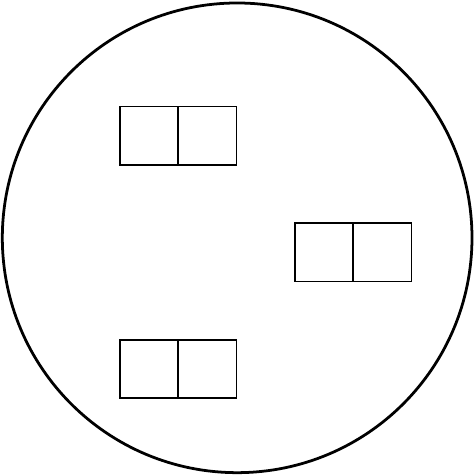}
\caption{$A_1$ theory on a sphere with three punctures }
\end{center}
\end{figure}

The notation introduced here for the correlator is done with a view towards the higher rank cases.  The $\mathfrak{sl_2}$ refers to the fact that Liouville CFT can be obtained from the $SL(2,\mathbb{R})$ WZW model under a gauging labeled by the principal embedding of $\mathfrak{sl_2} \rightarrow \mathfrak{sl_2}$ and the $[1^2]$ refers to the partition of $2=1+1$ that corresponds to the only non-trivial regular puncture coming from a codimension two defect of the $A_1$ theory\footnote{Going forward, the notation $V[\mathfrak{g}]_{g,[\ldots]}$ will be used to denote a correlator in the Toda theory labeled by a principal embedding of $\mathfrak{sl_2} \rightarrow \mathfrak{g}$ on a genus $g$ surface with punctures which are labeled by some representation theoretic data contained in the $[\ldots]$. }. The $\Lambda$ dependent factors that occur in the above formula follow from an analysis of scaling properties of Liouville correlators \cite{Knizhnik:1988ak,Distler:1988jt,David:1988hj}. The complete formula was proposed in  \cite{Dorn:1994xn,Zamolodchikov:1995aa}  along with some evidence for why this is true. It was then derived by Teschner using a recursion relation \cite{Teschner:1995yf}. 
Now, introduce a quantity that will be called the stripped correlator \footnote{An earlier draft of the paper was phrased in terms of scale factors for the original correlator and not the stripped correlator and this led to some inaccurate statements (for $g \neq 1$) in the calculation of scale factors via scaling behaviour of $\Upsilon$ function. While the original correlator also has scale factors, the one that is directly relevant for purposes of the AGT conjecture is the stripped one. I thank the anonymous referee for comments in this context.},
\begin{equation}
\hat{V}[\mathfrak{sl_2}]_{0,([1^2],[1^2],[1^2])} =  \frac{V[\mathfrak{sl_2}]_{0,([1^2],[1^2],[1^2])}}{\Upsilon(b) \Upsilon(2\alpha_1) \Upsilon(2\alpha_2) \Upsilon(2 \alpha_3)}
\end{equation}

It is the quantity $\hat{V}[\mathfrak{sl_2}]_{0,([1^2],[1^2],[1^2])}$ that seems most appropriate to identify as the partition function of four hypermultiplets. One expects that this quantity should posses an anomalous scaling term just like the one calculated in the previous section. And it indeed does have such an anomaly term and it matches exactly with that for a theory of four hypermultiplets ($n_h =4, n_v=0$).  This can be seen by noting the scaling behaviour of the $\Upsilon$ function (See Appendix B),

\begin{equation}
\Upsilon( \mu x ; \mu \epsilon_1, \mu  \epsilon_2) = \mu^{2 \zeta^B_2(0,x;\epsilon_1,\epsilon_2)} \Upsilon (x; \epsilon_1 ,\epsilon_2).  
\label{upsilonscaling}
\end{equation} 

There are a total of $\Upsilon(x)$ factors in the denominator of  $\hat{V}[\mathfrak{sl_2}]_{0,([1^2],[1^2],[1^2])}$ whose arguments take the value $x=1$ in the $m_i \rightarrow 0$ limit of $b=1$ Liouville theory. From Appendix B, note that $2\zeta^B_2(0,1;1,1)=-1/6$.  This implies (in the $m_i \rightarrow 0$ limit),
\begin{equation}
\hat{V}[\mathfrak{sl_2}]_{0,([1^2],[1^2],[1^2])} = \mu^{4/6} \hat{V}[\mathfrak{sl_2}]_{0,([1^2],[1^2],[1^2])}^{R_0=1}.
\label{threepointscaling}
\end{equation}

The factor $\mu^{4/6}$ matches with $\mu^{4a}$ for this theory and is thus in keeping with expectations. The dependence on the parameter $\mu = R_0^{-1}$  is usually suppressed when the Liouville correlators are analyzed. It had been additionally brought out here for it serves the useful purpose of encoding the Euler anomaly of the associated 4d SCFT which in this case is a trivial theory of four free hypermultiplets. For an exception on this matter, see \cite{Dorn:1991dr} where additional dimensionful parameters appear in the expression for the Liouville correlator $V[\mathfrak{sl_2}]_{0,([1^2],[1^2],[1^2])}$.  However, note that the exponent of the additional dimensionful parameter in \cite{Dorn:1991dr} is independent of the operator insertions. This wont be the true in what follows. The exponent of $\mu$ will have an important (and very subtle) dependence on the number and type of operator insertions. It turns out that for the case of the three point function, there is a plausible argument where the path integral description can be used to obtain the dependence on $\mu$. Consider,
\begin{equation}
V[\mathfrak{sl_2}]_{0,([1^2],[1^2],[1^2])} := \langle \mathcal{O}_1\mathcal{O}_2\mathcal{O}_3 \rangle = \int d [\phi]  e^{-S_{L,\text{disc}}} \prod_{i=1}^{3} e^{2 \alpha_i \phi}.
\label{3pointpi}
\end{equation}
Let us restrict ourselves to the case that corresponds to setting all the hypermultiplet masses $m_i$ to zero. Note that a primary operator $e^{2 \alpha_i \phi}$ modifies the boundary condition close to the insertion to $\phi = 2 \Re(\alpha) \log(r_i/R_0)$. To keep the action finite, one needs to introduce additional terms that are local to the punctures,
\begin{eqnarray*}
S_{L,\text{disc}} =  \frac{1}{72 \pi} \int_D \sqrt{\hat{g}} d^2 z \bigg ( \frac{1}{2} \hat{g}^{ab} \partial_a \phi \partial_b \phi   + 2 \pi \Lambda e^{2b\phi}  \bigg) &+& \frac{Q}{12 \pi}\int_{\partial D} \phi d \theta + \frac{1}{3} Q^2 \log (R/R_0) \\ &+& \sum_{i=1}^{3} \bigg( -   \frac{\Re(\alpha)}{6\pi}\int_{\partial c_i} \phi d \theta -  \frac{2}{3} \Re(\alpha)^2 \log (R/R_0) \bigg).
\label{actionreg}
\end{eqnarray*}
 For a translationally invariant measure $d [\phi]$, the $R_0$ contributions arise directly from the integrand. The global contribution is from the boundary term in $S_{cl}$ that is associated to the curvature insertion and is given by $(R_0)^{+Q^2/3}$. For the punctures, $\Re(\alpha_i) = Q/2$. So, each such operator insertion contributes $(R_0)^{-Q^2/6}$. Collecting these gives,
\begin{equation}
V[\mathfrak{sl_2}]_{0,([1^2],[1^2],[1^2])} = \mu^{Q^2/6} V[\mathfrak{sl_2}]_{0,([1^2],[1^2],[1^2])}^{R_0=1}.
\end{equation}
For the case of a round sphere, we have $Q=2$ and this implies $Q^2/6=2/3$. This is identified with the quantity $4a (= n_h/6)$  for a theory of four free hypermultiplets while $R_0$ is identified with the radius of the four sphere that was used as background for defining the partition function of the theory. Here, a comment on the unconventional normalization in $S_{L,disc}$ is required. The normalization of $\phi$ was chosen such that the dependence of $\mu$ for the three point function agrees with the corresponding value for $4a$. Equivalently, one could have picked the this factor such that the $n_h$ value for a single full puncture equals $4$. But, once it has been fixed, there are no free parameters. There will be similar choice of normalization later when the local contributions to these scale factors from are considered from a WZW point of view.

The calculation above reproduces the scale factor in \ref{threepointscaling}. When the scale factor is calculated from the $\Upsilon$ functions, the exact origin of the $\mu$ parameter is somewhat obscured by the regularization that is implicit in final form the DOZZ result. The path integral sheds \textit{some} light on how the scale factor enters into the picture via regularization. But, this is still incomplete since no such argument seems to be available readily for higher point functions. From (\ref{upsilonscaling}), it is also clear that the overall scale factor is sensitive to the analytical structure of the correlator. This relationship is most straightforward when  a correlator that corresponds to a free 4d theory is considered. In this case, the scale factor is purely from the $n_h$ contributions. The number of polar divisors in the correlation function is equal to $n_h$. In the example just considered, the number of polar divisors for the DOZZ three point function is $4$ and this indeed matches with the $n_h$ for a theory of four hypermultiplets.

A point worth emphasizing here is that the AGT primary map, namely the relation $\alpha_i = Q/2 + m_i$, is written after a dimensionful scale (the radius of the four sphere) is set to be unity. The goal of making the Euler anomaly explicit can alternatively be stated as that of making the dependence on this scale explicit in the correlators.

\subsection{Higher point functions}

Once the three point function is known, the higher point functions for Liouville can be obtained by the bootstrap procedure. This entails picking a factorization limit for the higher point function and writing the $n-$point function as an integral/sum over states in the $3g-3+n$ factorization channels with the integrand being built out of the $2g-2+n$ three point functions and appropriate conformal blocks. Confirming that the analytical structure of the resulting $n-$point functions is in keeping with the \textit{a priori} expectations (say, from a path integral point of view) involves a delicate interplay between the DOZZ three point function, the conformal block and the representation theory of the Virasoro algebra \cite{Teschner:2001rv}  (See Appendix \ref{bootstrap} for a short review). When there are enough punctures on both sides of the channel, the channel state is a primary with a momentum of the form $\alpha = Q/2 + i\mathbb{R}^{+}$ \cite{Polchinski:1990mh,Seiberg:1990eb}. The correlation functions built in the above fashion are also required to obey the generalized crossing relations. This imposes a highly nontrivial constraint on the three point function. For the case of Liouville, it has been checked that the DOZZ proposal does satisfy these constraints \cite{Ponsot:1999uf,Hadasz:2009sw}. Let us proceed now by looking at some examples of how the scale factor can be calculated for these higher point functions.
\subsubsection{$V[\mathfrak{sl_2}]_{0,([1^2],[1^2],[1^2],[1^2])}$}

\begin{figure}
\begin{center}
\includegraphics{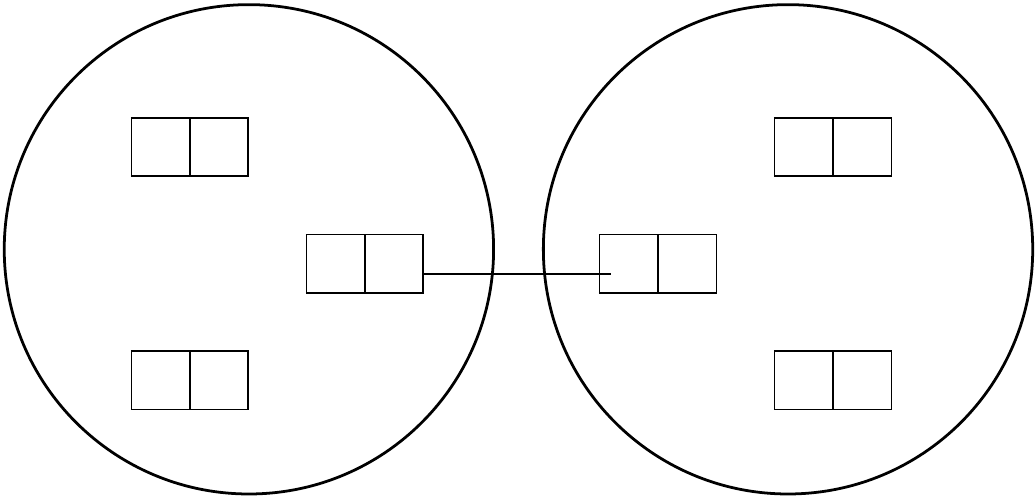}

\caption{$A_1$ theory on a sphere with four punctures in a degenerating limit. }
\end{center}
\end{figure}

This is the correlator corresponding to $\mathcal{N} =2$ SYM with gauge group $SU(2)$ and $N_f=4$. The flavor symmetry for this theory is $SO(8)$. The theory has four mass deformation parameters which can each be assigned to a $SU(2)$ flavor subgroup of $SO(8)$. These mass parameters will be related to the Liouville momenta in the following fashion
\begin{equation*}
 \alpha_i = \frac{Q}{2} + m_i.
\end{equation*}
The eigenvalues of the mass matrix are $m_1 + m_2$, $m_1 - m_2$, $m_3+ m_4$ and $m_3 - m_4$. 

To write down the four point function in Liouville theory, $\alpha_i, \alpha$ are initially taken to lie on the physical line. That is, $\alpha_i = Q/2 + i s_i^+, \alpha =Q/2 + is^+$ for $s_i^+, s^+ \in \mathbb{R}^{+}$. The four point function can then be written as
\begin{eqnarray*}
 Z_{S^4} = && V[\mathfrak{sl_2}]_{0,([1^2],[1^2],[1^2],[1^2])}(\alpha_1,\alpha_2,\alpha_3,\alpha_4) = \\ &&\int_{\alpha \in {\frac{Q}{2}+ i s^+}}d \alpha C(\alpha_1, \alpha_2, \alpha)  C (Q - \alpha,\alpha_3,\alpha_4) \mathcal{F}_{12}^{34} (c, \Delta_{\alpha}, z_i) \mathcal{F}_{12}^{34}(c, \Delta_{Q -\alpha}, \bar{z}_i).
\label{fourpoint}
\end{eqnarray*}
In writing this, the fact that when $\alpha \in {\frac{Q}{2}+ i s}$, $\bar{\alpha}=Q-\alpha$ has been used. Now, using the symmetry of the entire integrand under the Weyl reflection $\alpha \rightarrow Q -\alpha $, one can unfold the integral to one over $\mathbb{R}$. This gives 
\begin{eqnarray*}
 V[\mathfrak{sl_2}]_{0,([1^2],[1^2],[1^2],[1^2])}&&(\alpha_1,\alpha_2,\alpha_3,\alpha_4) = \\ &&\frac{1}{2} \int_{\alpha \in {\frac{Q}{2}+ i s}}d \alpha C(\alpha_1, \alpha_2, \alpha)  C (Q - \alpha,\alpha_3,\alpha_4) \mathcal{F}_{12}^{34} (c, \Delta_{\alpha}, z_i) \mathcal{F}_{12}^{34}(c, \Delta_{Q -\alpha}, \bar{z}_i),
\end{eqnarray*}
where $s \in \mathbb{R}$.
As with the three point function, let us defined the stripped four point function,
\begin{equation}
 \hat{V}[\mathfrak{sl_2}]_{0,([1^2],[1^2],[1^2],[1^2])} = \frac{ V[\mathfrak{sl_2}]_{0,([1^2],[1^2],[1^2],[1^2])}}{\Upsilon(b) \Upsilon(2 \alpha_1)\Upsilon(2 \alpha_2)\Upsilon(2 \alpha_3)\Upsilon(2 \alpha_4)}
\end{equation}
To calculate the overall $R_0$ dependence, the anomalous terms from the $\Upsilon$ factors should be collected. A simple variable change collects the extra factors from the integration over channel momenta and the conformal blocks. The contribution from the eight polar divisors in the integrand is also straightforward to calculate and is equivalent to the contribution from the denominator in \ref{oneloop}. As for the term $\Upsilon (2\alpha) \Upsilon ( 2Q-2\alpha)$, this can be rewritten terms of the $H$ function in order to make the Vandermonde factor explicit (as in \cite{Alday:2009aq}). Let us note here the steps involved,
\begin{eqnarray}
\Upsilon (2\alpha) \Upsilon ( 2Q-2\alpha) &=& \Upsilon (Q + 2ia) \Upsilon (Q- 2 ia) \\
	&=& \frac{1}{\Gamma_2 (Q+2ia) \Gamma_2(-2 ia)}\frac{1}{\Gamma_2(Q-2ia) \Gamma_2 (2 ia)}
\end{eqnarray}
Recalling the following property (Appendix B) of the digamma function,
\begin{equation}
[\Gamma_2 (x + \epsilon_1 + \epsilon_2) \Gamma_2(x)]^{-1} = x [\Gamma_2(x+\epsilon_1)\Gamma_2(x+\epsilon_2)]
\end{equation}
and applying it to case of $\epsilon_1=b,\epsilon_2=1/b$, 
\begin{eqnarray}
\Upsilon (2\alpha) \Upsilon ( 2Q-2\alpha) &=& (2 ia)^2 [\Gamma_2(b+2 ia) \Gamma_2 (b^{-1}+ 2 ia)]^{-1} [\Gamma_2(b-2ia)\Gamma_2(b^{-1}-2ia)] \nonumber \\ 
&=& -4 a^2 H(2 ia) H(- 2 ia).
\end{eqnarray}
The above factor taken together with the single $\Upsilon(b)$ that remains in $\hat{V}[\mathfrak{sl_2}]_{0,([1^2],[1^2],[1^2],[1^2])}$ provide the numerator in the expression for $Z_{1-loop}$ (\ref{oneloop}) together with Vandermonde factor. The calculation of the scale factor is thus reduced the calculation that we already performed.  So, we have 
\begin{equation}
\hat{V}[\mathfrak{sl_2}]_{0,([1^2],[1^2],[1^2],[1^2])} = \mu^{23/6} \hat{V}[\mathfrak{sl_2}]_{0,([1^2],[1^2],[1^2],[1^2])}^{R_0=1}. 
\end{equation}
The exponent of $R_0$ can be interpreted as $4a$ and this indeed matches \ref{eulercharge} for $N=2$.
\subsubsection{$V[\mathfrak{sl_2}]_{1,([1^2])}$}
\begin{figure}
\begin{center}
\includegraphics{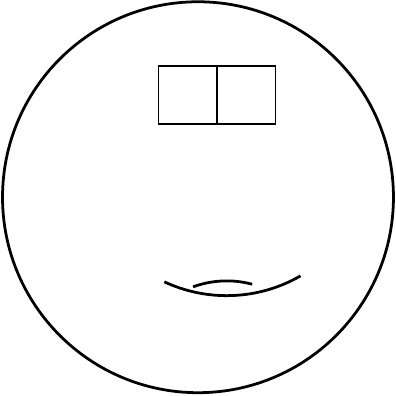}
\caption{$A_1$ theory on a torus with one puncture }
\end{center}
\end{figure}
 For an arbitrary mass deformation, this theory corresponds to $\mathcal{N} = 2^*$ SYM with $SU(2)$ gauge group with a hypermultiplet in the adjoint and one free hypermultiplet. The corresponding Liouville correlator can be expressed in terms of the one point conformal block for the torus.
\begin{equation*}
 V[\mathfrak{sl_2}]_{1,([1^2])} (\alpha_1) = \int_{\alpha \in Q/2 + s } d \alpha  C(Q-\alpha,\alpha_1,\alpha) \mathcal{F}_{\alpha_1}(\Delta_{\alpha},q) \mathcal{F}_{\alpha_1}(\Delta_{Q-\alpha},\bar{q}).
\end{equation*}  
The stripped correlator in the $g=1$ case is defined as
\begin{equation}
 \hat{V}[\mathfrak{sl_2}]_{1,([1^2])} (\alpha_1) =  \frac{V[\mathfrak{sl_2}]_{1,([1^2])} (\alpha_1)}{\Upsilon(2\alpha_1)}.
\end{equation}
Calculating the $R_0$ dependence as in the case of the four point function,
\begin{equation}
\hat{V}[\mathfrak{sl_2}]_{1,([1^2])} = \mu^{19/6} \hat{V}[\mathfrak{sl_2}]_{1,([1^2])}^{R_0=1}.
\end{equation}
Ignoring the contribution of a decoupled hypermultiplet(with $4a = 1/6$) gives the expected answer that $4a=3$ for the $N=2^*$ theory. 

For higher point functions on arbitrary surfaces, one proceeds in a similar manner by defining the stripped correlator as
\begin{equation}
\hat{V}_{g,[\ldots]} (\alpha_1, \alpha_2 \ldots \alpha_n) = \frac{V_{g,[\ldots]} (\alpha_1, \alpha_2 \ldots \alpha_n) \Upsilon(b)^{g-1}}{\prod_i \Upsilon(2 \alpha_i)},
\label{NpointScale}
\end{equation}
where $V_{g,[\ldots]} (\alpha_1, \alpha_2 \ldots \alpha_n)$ is the Liouville correlator built out of $(2g-2+n)$ DOZZ three point functions and $(3g-3+n)$ factorizing channels. Calculating the contributions to the scale factor directly from \ref{NpointScale},
\begin{eqnarray}
4a &=& (2g -2+n) \bigg( 3\frac{5}{6} - \frac{1}{6}  + 4 \frac{1}{6}\bigg) + (3g-3+n) - \frac{5}{6} n - \frac{1}{6} (g-1) \\
 &=& \frac{53}{6} (g-1) + \frac{19n}{6}.
\end{eqnarray}

From (\ref{NhandNv}), $n_h = 8(g-1) + 4n$, $n_v=9(g-1) + 3n$ and one sees immediately that $4a$ calculated above satisfies
\begin{equation}
4a = n_v + \frac{n_h - n_v}{6}.
\end{equation}

\subsection{Liouville theory from a gauged WZW perspective}
\label{liouvillewzw}

Before proceeding to discuss the higher rank generalizations, it is useful to recast the scale factor calculations in an alternate language. It is well known that classical Liouville theory can be obtained via a Hamiltonian reduction starting from the $SL(2,\mathbb{R})$ WZW model. A quantum version of this reduction (which has been the subject of a rich variety of studies from various different points of view. See for instance \cite{feigin1990representations}, \cite{dotsenko1990free,furlan1993solutions} and  \cite{Ribault:2005wp,Hikida:2007tq,Ribault:2008si,Giribet:2008ix}) is then expected to yield Liouville conformal field theory. This point of view is powerful since it permits an easy generalization to higher rank cases where a non rational CFT with W-symmetry is obtained for every inequivalent (upto $\mathfrak{g}$ conjugacy) $\sigma : \mathfrak{sl_2} \rightarrow \mathfrak{sl}_N$ . In the $\mathfrak{g}=\mathfrak{sl_2}$ case considered here, the only non-trivial embedding is the principal embedding and this corresponds to Liouville CFT. With this in mind, let us look at how the spectrum of primaries in Liouville can be related to a set of WZW primaries. In its Wakimoto realization, this model is realized in terms of a scalar field $\phi$ and bosonic ghosts $\beta,\gamma$ with the following bosonization rules
\begin{eqnarray}
J^+ &=& -\beta(z) \gamma(z)^2 +  \alpha \gamma(z) \partial \phi(z) + k \partial \gamma(z), \\
J^3(z) &=& \beta(z) \gamma(z) - \frac{ \alpha}{2} \partial \phi(z), \\
J^-(z) &=& \beta(z).
\end{eqnarray}
with $\alpha^2 = 2k-4$.  Now, consider the primary field $P(j)$  whose free field realization is $\gamma^{-j} \overline{\gamma}^{-j}e^{(j+1)\phi}$. This operator is identified with a  Liouville primary of the form $e^{\alpha \phi}$ (upto a scale that will be fixed momentarily) where $\alpha = -j b$ . Naively, the conformal dimensions of the primaries match. That is,  
\begin{equation}
\Delta_\alpha = \Delta_j - j,
\label{momentumspin}
\end{equation}
where $\Delta_\alpha = \alpha (Q - \alpha)$ and $\Delta_j = -\frac{j (j+1)}{k-2}$ with the identification $b^2 = k-2$. In early investigations of these gauged WZW models, it was shown that the two and three point functions of Liouville can be obtained exactly under the above identification of primaries along with (\ref{momentumspin}) holding \cite{dotsenko1990free,furlan1993solutions}.

 One of the advantages of the WZW prescription is that the classical solutions are perfectly regular. In the WZW language, there is no singularity in the local solution near the insertion of the puncture and consequently, there are no regularizing terms in the classical action. So, where does the dependence of $R_0$ arise ?  I argue that it arises from carefully considering the dimensionful factors that enter in the relationship between the Liouville and WZW primaries. First, in the gauged WZW model, one works with an 'improved' stress energy tensor 
\begin{equation}
\hat{T}(z) = T(z)_{WZW} - \partial J^{3}(z),
\end{equation}
so that the constraint $J^- =1$ can be imposed without breaking conformal invariance. Under the improved stress energy tensor, the primary $P(j)$ has a shifted dimension $\hat{\Delta}=\Delta_j - j - j$ . To keep the map between primaries intact along with relation $\Delta_\alpha = \Delta_j - j $, a scale factor that offsets the shift in dimension of $P(j)$ should be included
\begin{equation}
e^{\alpha \phi} \equiv (R/R_0)^{+j}\gamma^{-j} \overline{\gamma}^{-j}e^{(j+1)\phi}.
\end{equation}
A further redefinition of $\phi$ is needed in order to match the normalization used in the previous section. It is chosen such that $j=-2(= -4(\rho,\rho))$ corresponds to the full puncture for a $b=1$ theory with $n_h/6=-2/3$. In this normalization,
\begin{equation}
e^{\alpha \tilde{\phi}} \equiv (R/R_0)^{+j/3}\gamma^{-j/3} \overline{\gamma}^{-j/3}e^{(j+1)\tilde{\phi}}.
\end{equation}
\section{Scale factors in Toda correlators I : Primaries and free theories}
\label{Antheories1}
In the Toda case, the WZW approach is much more convenient to capture the local $n_h$ contributions to the scale factor since a Toda action with appropriate boundary terms is not readily available for an arbitrary codimension two defect. However, the global $n_h$ contribution will always be computed using the curvature insertion in the Toda action. This asymmetric treatment is purely one of convenience. A complete understanding of boundary actions in Toda theory might be a way to obtain a more uniform treatment \cite{Fateev:2010za}. 

The most general Toda theory of type $A$ can be obtained by a gauging of the $SL(n,\mathbb{R})$ WZW model. Unlike the case of $A_1$, the higher rank cases offer more than one ways of gauging the $SL(n,\mathbb{R})_L \times SL(n,\mathbb{R})_R$ symmetry such that conformal invariance in preserved\cite{Bershadsky:1989mf,Bershadsky:1990bg}.  An optimal way to index inequivalent Toda theories is by associating a $\mathfrak{sl_2}$ embedding $\sigma :\mathfrak{sl_2} \rightarrow \mathfrak{sl}_N $ for every such gauging \cite{Feher:1992ed,deBoer:1993iz}. Each of the theories obtained by a nontrivial embedding $\sigma$ has a $\mathcal{W}-$ symmetry whose chiral algebra is called a $\mathcal{W}-$ algebra. This algebra is a non-linear extension of the Virasoro symmetry by currents $\{\mathbf{W}^i(z)\}$  of spin $ i (> 2)$. The unique spin 2 current in the chiral algebra is the  stress energy tensor $\mathbf{T}(z)\equiv\mathbf{W}^2(z)$.

 As with $Sl(2,\mathbb{R})$, consider the Wakimoto realization of the $SL(n,\mathbb{R})$ model with the required number of $\beta, \gamma,\phi$ fields.  The following constraints are imposed \cite{Feher:1992ed}
\begin{eqnarray}
J(x) &=& K e + j(x),\\
\tilde{J}(x) &=& - K f + \tilde{j}(x).
\end{eqnarray}
where ${e,f,h}$ are the images of the standard $\mathfrak{sl_2}$ generators and  $ j(x) \in \mathfrak{g}_{\ge0}$ and $\tilde{j} \in \mathfrak{g}_{\le0}$ \footnote{$K$ is a potentially dimensionful constant.}.

When the grading is even, the system of constraints is first class. When the grading has odd pieces, then at first sight, the system is not first class. One can introduce auxiliary fields (as in \cite{Bershadsky:1990bg}) or consider a grading by a different element $M$ such that $[M,h]=0,[M,e]=2e,[M,f]=-2f$ \cite{Feher:1992ed}. In the latter case, it is possible to define a new set of constraints (now first class) equivalent to the original.

 In this paper, \text{only} the theories obtained by the principal embedding will be considered. It has the following action on the disc (written in the same unconventional normalization that was used for the Liouville case),
\begin{equation}
S_{T,\text{disc}} =  \frac{1}{72 \pi} \int_D \sqrt{\hat{g}} d^2 z \bigg ( \frac{1}{2} \hat{g}^{ab} \partial_a \phi \partial_b \phi   + \sum_{i=1}^{n-1} 2 \pi \Lambda e^{2b(e_i,\phi)}  \bigg) + \frac{1}{6\pi}\int_{\partial D} (Q,\phi) d \theta + \frac{2}{3} (Q,Q) \log (R/R_0).
\label{actionTodaN}
\end{equation}
The conformal transformations that leave the above action invariant (classically) are
\begin{eqnarray*}
z' &=& w(z), \\
\phi' &=& \phi - Q \rho \log \bigg( \frac{\partial w}{\partial z}\bigg)^2 ,
\end{eqnarray*}
and the field $\phi$ obeys the following boundary condition at the boundary of the disc
\begin{equation}
\phi = -Q \rho \log (R/R_0)  + \mathcal{O} (1).
\end{equation}
The chiral algebra for this theory is generated by the currents $\{\mathbf{W}^i(z)\}$  of spin $ i = 2 \ldots n-1$. The spins of the currents are identified with the exponents of the group $SL(n,\mathbb{R})$.  The global $n_h$ contribution arises from the boundary term due to the curvature insertion (specializing to $Q=2$ and generalizing the relevant boundary term for a surface of arbitrary genus), we get the $n_h$ dependent contribution to $4a^{global}$,
\begin{equation}
(4 a)^{global}_{n_h} = \frac{8}{3} (g-1) (\rho,\rho).
\end{equation}
Now, using $(4 a)^{global}_{n_h} = n_h/6$), it follows that
\begin{equation}
n_h^{global} = 16 (g-1) (\rho,\rho).
\end{equation}
This matches with (\ref{NhNvformulae}) once we use the Freudenthal-de Vries formula for $(\rho,\rho)$. We will now proceed to analyze an interesting family of primary operators also indexed by inequivalent embeddings of $\rho : \mathfrak{sl_2} \rightarrow \mathfrak{g}$. In type $A$, the identification of these primaries has been done in \cite{Kanno:2009ga}. Following \cite{Kanno:2009ga}, these states are referred to as semi-degenerate primaries. They will be related to certain primaries in the WZW model. To go beyond just calculating the $n_h$ contributions, it will also be useful to associate an irreducible representation of the Weyl group to each of those operators. 

\subsection{Toda primaries from a gauged WZW perspective}

The set of semi-degenerate primaries relevant for the AGT correspondence was constructed in \cite{Kanno:2009ga} by applying the screening operators $\mathcal{S}_i^{(\pm)}$ to Toda primary whose momentum satisfies certain conditions.  The screening operators have the following form
\begin{equation}
\mathcal{S}_i^{(\pm)} = \int \frac{dz}{2 \pi i} e^{(\beta e_i \cdot \phi)},
\end{equation}
where $e_i$ are the simple co-roots of $\mathfrak{sl}_N$. Requiring that these operators have $\Delta=1$ forces $\beta$ to be either $\beta_+ = -b$ or $\beta_- = -1/b$. The screening operators have the special property that they commute with the generators of the $\mathcal{W}$ algebra. That is $[W^k_l,S_i^{\pm}]=0$. Now, the state $(\mathcal{S}^{\pm})^{n^{\pm}} | \alpha - n_{\pm} \beta_{\pm} e_i>$ either vanishes identically or has a null state at level $n_+ n_-$. For the latter to happen, the $\alpha$ have to satisfy
\begin{equation}
(\alpha,e_i) = (1- n_j^+)\alpha_+ + (1- n_j^-)\alpha_-,
\end{equation}
for some $j$. If the null vectors are taken to appear at level one, the above condition is simplified to
\begin{equation}
(\alpha,e_i) = 0,
\label{NullCond}
\end{equation}
for ${e_i}$ being some subset of simple co-roots. Having recalled the construction in \cite{Kanno:2009ga}, we proceed to obtain these primaries in the gauged WZW setting. The proposed map is the following
\begin{equation}
e^{(\alpha, \phi)} \equiv (R/R_0)^{8(\rho,\rho)-4(\rho,h)+\frac{1}{2}\text{dim}\mathfrak{g}^h_{1}} \prod_i \gamma_i \prod_k \bar{\gamma_k} \times e^{(j+2 \rho ,\phi)},
\label{primarymap}
\end{equation} 
for some specific choices of $\alpha$ (and consequently of $j$). The different semi-degenerate states are obtained for the choices of $\alpha$ outlined in \cite{Kanno:2009ga}. For the case $b=1$, the set in \cite{Kanno:2009ga} can be obtained by setting $\alpha = 2\rho -\lambda$ where $\lambda$ is twice the Weyl vector of a subalgebra of $\mathfrak{sl}_N$. The spin $j$ in the WZW primary is obtained using $j=-\alpha$. The justification for the scale factor in the above map is similar in spirit to the one  encountered in the case of Liouville (see Section \ref{liouvillewzw} ) but the details are complicated by the  wider variety of semi-degenerate state that are available in the higher rank Toda theories. This requires the introduction of some representation theoretic notions.

First, note that considerations of scaling in Toda theory involve more possibilities in that one has to first pick a weight vector and consider scaling in the direction of that weight vector. The maximal puncture is the one that is not invariant under a scaling along any weight vector. In other words, for a maximal puncture, there is no $\lambda \in \Lambda$ such that $(\alpha, \lambda)=0$. For other smaller punctures, there always exists such a $\lambda$ and the 'smallness' of the puncture is related to how 'big' the $\lambda$ is. The scare quotes are included to highlight that this notion of small/big is not rigorous since two sets (the set of regular punctures and the set of weight vectors) admit only a poset structure and it may turn out that certain pairs do not have an order relationship.  The $h$ in the above formula is obtained in the following way. Take the subalgebra $\mathfrak{l}$ of $\mathfrak{sl}_N$ for which $\lambda$ is twice the Weyl vector ($2\rho_\mathfrak{l}$). Let ${e_i}$ be a set of simple co-roots for this subalgebra. Impose the null vector conditions \ref{NullCond} for this set.  Now, consider orbit of $\lambda$ under $W[\mathfrak{sl}_N]$. There is a unique element $h = w \lambda$ for $w \in W[\mathfrak{sl}_N]$ and $h \in \Lambda^+$, the set of dominant weights of $\mathfrak{g}$. This dominant weight is the Dynkin element (See Appendix for explanation of this terminology) of a nilpotent orbit in $\mathfrak{sl}_N$. Such orbits are classified by partitions of $N$. One can translate between the different quantities in the following way. Given a partition $[n_1 n_2 \ldots n_k]$ such that $\sum n_i = N$, write $\lambda$ as $(-n_1 +1, -n_1 +3, \ldots n_1 - 1, -n_2 +1, \ldots n_2 - 1 \cdots -n_k +1, \ldots n_k - 1)$. Reordering the elements of $\lambda$ such that they are non-decreasing gives us $h$, the Dynkin element. 

The element $h$ occurs as the semi simple element in the $\mathfrak{sl_2}$ triple $\{ e,f,h\}$ associated to the corresponding embedding. The lie algebra $\mathfrak{g}$ has a natural grading defined by the $h$ eigenvalue
\begin{equation}
\mathfrak{g} = \oplus_j {\mathfrak{g}_i} = \oplus_{j<0} {\mathfrak{g}_i} +  {\mathfrak{g}_0}+  \oplus_{i>0} {\mathfrak{g}_i}.
\end{equation}

We can now turn to the interpretation of the scale factor in \ref{primarymap}. Consider the special case : $j$ such that $h$ is trivial ($\lambda =0$). This corresponds to a 'maximal' puncture.  As with the case of Liouville, the necessity of using a modified stress tensor $\hat{T}^\rho(z)$ ($\rho$ denotes the fact that this is the stress tensor for the principal Toda theory) introduces extra contributions to the scaling dimension. To avoid spoiling the relationship $\Delta_\alpha = \Delta_j - (j,2\rho)$, there is a need to introduce a scale factor of the form $ (R/R_0)^{4(\rho,\rho)}$. When $h$ is non trivial, there are some scalings for which the primary is invariant (as opposed to transforming by a scale factor). Local to the primary insertion, associate a $\mathfrak{sl_2}$ embedding with Dynkin element $h$ and consider the spectrum of $\gamma$ fields associated to this grading. Their dimensions are given by how they behave under a scaling defined by $\hat{T}^h(z)$. If the embedding is even ($dim \mathfrak{g}_{\pm i} = 0$ for i odd), one would like to remove the contribution to the scaling dimension from the corresponding set of $\gamma$ fields. When the embedding is not even, this procedure will work if a grading under a different element $M$ is considered. This $M$ is such that it provides an even grading while obeying $[e,M] = 2e, [f,M] = -2f,[h,M]=0$ \cite{Feher:1992ed}. Under the new grading, the dimension of  $\mathfrak{g}_{\geq 2}$ increases by $\frac{1}{2} \text{dim} \mathfrak{g}_{1}$. So, a full accounting of the dimensional factors produces the exponent of $R/R_0$ in \ref{primarymap}.
As with the Liouville case, $\phi$ needs to be normalized such that $h=0$ produces the correct $n_h$ contribution from a full puncture. In this normalization,
\begin{equation}
e^{(\alpha, \phi)} \equiv (R/R_0)^{\frac{8(\rho,\rho)-4(\rho,h)+\frac{1}{2}\text{dim}\mathfrak{g_{1}}}{6}} \prod_i \gamma_i \prod_k \bar{\gamma_k} \times e^{(j+2 \rho ,\phi)}.
\label{primarymap2}
\end{equation} 

The exponent of $R_0$ is recognized as the local contribution to $n_h/6$ from CDT \cite{Chacaltana:2012zy}\footnote{A clarification regarding the notation is in order. What is called $\text{dim} \mathfrak{g}_1$ here is the same as $\text{dim} \mathfrak{g}_{1/2}$ of \cite{Chacaltana:2012zy}. The difference in notation arises from the choice of normalization of $h$.}. One would like to believe that the other local properties ascribed to this class of codimension two defects of the six dimensional theories should also have a description in terms of properties of the corresponding semi-degenerate operators in Toda theory. In order for this dictionary to be built further, it is important to associate to every semi-degenerate primary a unique irrep of the Weyl group.

\subsection{Toda primaries and representations of Weyl groups}

In this section, a representation of the Weyl group $W[\mathfrak{sl}_N]=S_N$ will be associated to every semi-degenerate primary in an $A_n$ Toda theory. Recall from the previous section that the momentum of a general semi-degenerate primary obeys $(\alpha,e_i)=0$ for $i=1\ldots k$. The ${e_i}$ are a subset of the set of simple co-roots $\Pi$. In the current case, they form a subsystem\footnote{More accurately, a conjugacy class of subsystems.}. Denote this set by $S_{N}$. Denote by $S_N^+$, the set of positive root of this subsystem. Let $\Lambda^+$ be the set of positive roots for $\mathfrak{g}$. Note here that when $h$ is zero, $S_N^+$ is empty and when $h$ is the Dynkin element of the principal nilpotent orbit, $S^+$ is $\Lambda^+$ .

 Using this data, one can obtain a unique irreducible representation of the Weyl group by a construction due to MacDonald \cite{macdonald1972some} \footnote{See the text \cite{carter1985finite} for an elaborate discussion of this construction and its generalization due to Lusztig and Lusztig-Spaltenstein.}. The co-root system lives naturally in $\mathfrak{h}$. Each co-root can be thought of as a linear functional on $\mathfrak{h}^*$. Now, construct the following rational polynomial on  $\mathfrak{h}^*$,
\begin{equation}
\pi = \prod_{e_\alpha \in  S_N^+} e_\alpha.
\end{equation}
Using this, construct a subalgebra of the symmetric algebra($\mathfrak{S}$) on $\mathfrak{h}^*$ by considering all polynomials $\mathcal{P} = w \pi$. This subalgebra is a $W-$ module and in fact, furnishes an irreducible representation of the Weyl group. 

It turns out that $\textit{all}$ irreps for Weyl groups of types $A,B,C$ can be obtained by considering the various inequivalent subsystems.

 The contribution to the total Coulomb branch dimension of the four dimensional theory from a primary that is labeled by a Nahm orbit $\mathcal{O}_N$ is actually related to the dimension of a dual orbit \cite{Chacaltana:2012zy}. This formula can be rewritten in terms of the cardinality of the set $S_n^+$ in the following way
\begin{equation}
d = |\Delta^+ | - |\Delta_{S_N}^+| = \frac{1}{2} \text{dim} \mathcal{O}_{P^t}.
\end{equation}
where $P$ is the partition type associated to the Nahm orbit $\mathcal{O}_N$ and $P^t$ is the transpose partition. 
Let $\phi_i$ be the generators of the full symmetric algebra. Let us additionally note here the formula 
\begin{equation}
n_v = \sum_i [2 \text{deg} (\phi_i) -1] - \sum_{h > 0}[2 h - 1] = 2 (2 \rho, 2 \rho - h) + \frac{1}{2} (\text{rank} \mathfrak{g} - \text{dim} \mathfrak{g}^h_0).
\label{nvlocal}
\end{equation}
This quantity is called $n_v$ since it will turn out to be the contribution of the codimension two defect to the effective number of vector multiplets. To give a flavor for the values $n_h,n_v$ in the various cases, the properties of Toda semi-degenerate states for the $A_2$, $A_3$ theories in are collected in Tables \ref{A2states} and \ref{A3states}. In the tables, the fundamental weights are denoted by $\omega_i$ and the nomenclature of a 'Nahm Orbit' and a 'Hitchin Orbit' is borrowed from \cite{Chacaltana:2012zy} and is in anticipation of the next Section.

\begin{center}
\begin{table}
\begin{center}
\begin{tabular}{|c|c|c|c|c|c|}
\hline 
$h$ & Nahm Orbit & Hitchin Orbit & Toda momentum($\alpha$) & $n_h$ & $n_v$ \\ 
\hline 
$(0,0,0)$&$[1^3]$ & $[3]$ & $2(\omega_1 + \omega_2)$ & $16$ & $13$ \\ 
\hline 
$(1,0,-1)$&$[2,1]$ & $[2,1]$ & $3\omega_1$ & $9$ & $8$ \\ 
\hline 
$(2,0,-2)$&$[3]$ & $[1^3]$ & $0$ & $0$ & $0$ \\ 
\hline 
\end{tabular} 
\caption{Semi-degenerate states in $A_2$ Toda theory.}
\end{center}
\label{A2states}
\end{table}
\end{center}

\begin{center}
\begin{table}
\begin{center}
\begin{tabular}{|c|c|c|c|c|c|}
\hline 
$h$& Nahm Orbit & Hitchin Orbit & Toda momentum($\alpha$) & $n_h$ & $n_v$ \\ 
\hline 
$(0,0,0,0)$&$[1^4]$ & $[4]$ & $2(\omega_1 + \omega_2 + \omega_3 )$ & $40$ & $34$ \\ 
\hline 
$(1,0,0,-1)$&$[2,1^2]$ & $[3,1]$ & $3\omega_2 + 2 \omega_1$ & $30$ & $27$ \\ 
\hline 
$(1,1,-1,-1)$&$[2,2]$ & $[2,2]$ & $4\omega_2$ & $24$ & $22$ \\
\hline
$(2,0,0,-1)$&$[3,1]$ & $[2,1^2]$ & $4\omega_1$ & $16$ & $15$ \\
\hline
$(3,1,-1,-3)$&$[4]$ & $[1^4]$ & $0$ & $0$ & $0$ \\
\hline 
\end{tabular} 
\caption{Semi-degenerate states in $A_3$ Toda theory.}
\end{center}
\label{A3states}
\end{table}
\end{center}

\subsection{Physical interpretation of the Lusztig-Spaltenstein map}

The appearance of the dimension formula for the dual orbit leads to an obvious question. Can the Lusztig-Spaltenstein map \footnote{As in \cite{Chacaltana:2012zy}, we will always refer to the modified Lusztig-Spaltenstein map as defined by Barbasch-Vogan. In the mathematical literature, it is often denoted as $d_{BV}$.} be obtained from the data just considered ? Recall here that this is a map from nilpotent orbits in $\mathfrak{g}$ to nilpotent orbits in the Langlands dual algebra $\mathfrak{g}^L$,
\begin{equation}
d_{LS} :  \rho \rightarrow \rho',
\end{equation}
where $\rho : \mathfrak{sl_2} \rightarrow \mathfrak{g} $ and $\rho' : \mathfrak{sl_2} \rightarrow \mathfrak{g}^L$. 
 
In the CDT description \cite{Chacaltana:2012zy} of this class of codimension two defects, a pair $(\rho,\rho')$ plays a central role\footnote{In cases outside of type $A$, there is also a discrete group.} in the description of a \textit{single} defect. In denoting $\rho$ as the 'Nahm data' and $\rho'$ as the 'Hitchin data', I have borrowed the terminology from \cite{Chacaltana:2012zy}.  In the setup here, the $h$ from the previous sections is the 'Nahm Data'. The 'Hitchin data', I propose, can be related to the 'Nahm data' using the irrep of the Weyl group constructed in the previous section. Crucial to this proposal is the existence of another way of constructing Weyl group representations due to Springer.  Springer showed that the Weyl group acts on the cohomology $H^*(\mathcal{B}_{e'},\mathbb{C})$ of the Borel variety fixed by the nilpotent orbit through $e'$.  Consider the  resolution of the nilpotent cone $\mathcal{N}$, \\

\begin{center}
\begin{tikzpicture}[node distance=2cm, auto]
  \node (P) {$T^*\mathcal{\mathcal{B}}\simeq \hat{\mathcal{N}}$};
    \node (BB) [left of=P] {$\mathcal{B}$};
  \node (A) [below of=P] {$\mathcal{N}$};
  \draw[->] (P) to node [swap] {$\mu$} (A);
  \draw[->] (P) to node {$\pi$} (BB);
\end{tikzpicture}
\end{center}
where $\hat{\mathcal{N}}$ is the space of pairs $\{(e,b)\mid b \in \mathcal{B}, e \in \mathcal{B} \cap \mathcal{N} \} $.

If we pick an element in $\mathcal{N}$ that is a representative of a nilpotent orbit, then the Springer fiber at that point is the Borel variety fixed by that nilpotent orbit. When $e'$ is in the zero orbit, for example, the Springer fiber is the full Borel variety $\mathcal{B}\cong G/B$ and the cohomology ring is identified with the co-invariant algebra (a result that is originally due to Borel \cite{borel1953cohomologie}).

The Springer map for type $A$ is obtained by identifying the irrep that occurs in the top degree of the cohomology. For other types, the top degree in general carries an irrep of $W \times A(e')$ where $A(e')$ is the component group of the centralizer of the nilpotent element. Since $A(e')$ is always trivial for type $A$, identifying the irreps of the Weyl group is straightforward.  For our current purposes, we will only need to know which Weyl group representation is assigned to a particular nilpotent orbit by the Springer map. This is available in the standard texts like \cite{carter1985finite} whose conventions are followed closely\footnote{To help with notation, note that the Nahm partition associated to $h$ here is the $\alpha^*$ partition in Prop. 11.4.1 of \cite{carter1985finite}.}. See also \cite{mcgovern2002adjoint,humphreys2011conjugacy} for introductions to Springer theory. Using that data, the  picture in Figure \ref{LSmap} is constructed.

\begin{center}
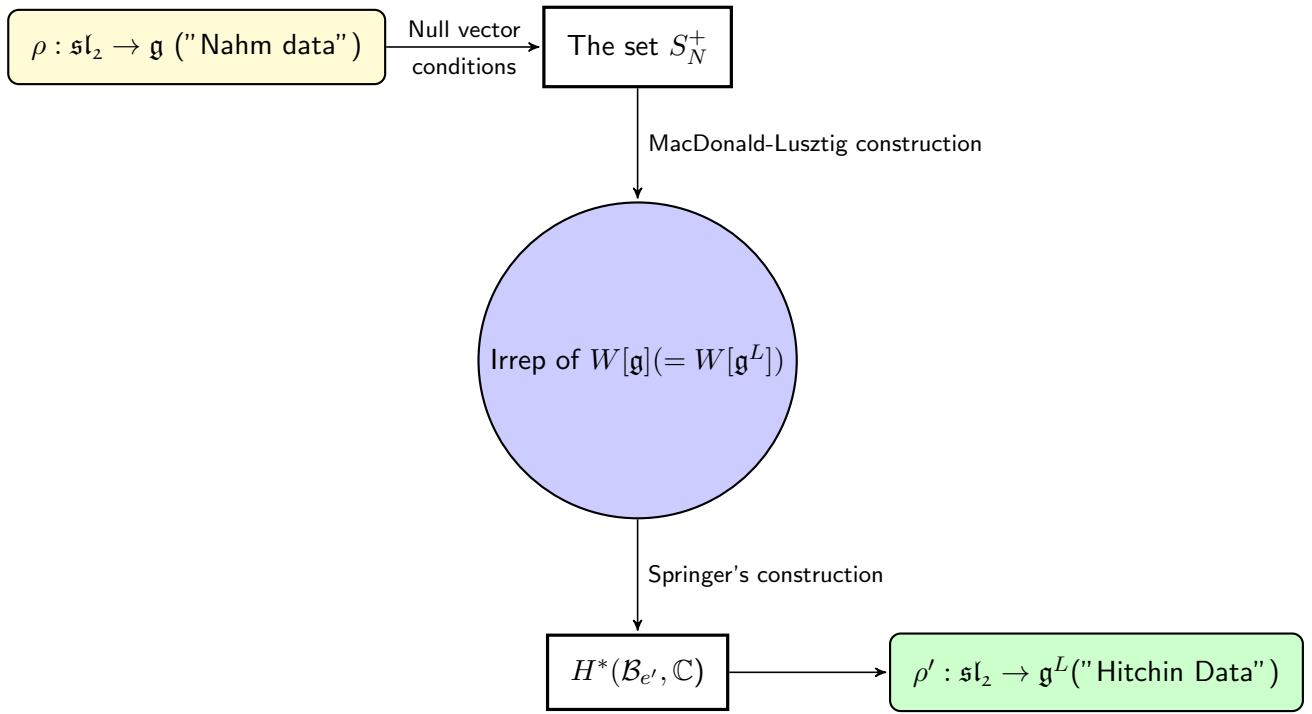
\begin{figure}[!h]
\label{proposal}
\begin{tikzpicture}[
  font=\sffamily,
  every matrix/.style={ampersand replacement=\&,column sep=1.2cm,row sep=1.5cm},
  source/.style={draw,thick,rounded corners,fill=yellow!20,inner sep=.3cm},
  process/.style={draw,thick,circle,fill=blue!20},
  sink/.style={source,fill=green!20},
  datastore/.style={draw,very thick,shape=rectangle,inner sep=.3cm},
  dots/.style={gray,scale=2},
  to/.style={->,>=stealth',shorten >=1pt,semithick,font=\sffamily\footnotesize},
  every node/.style={align=center}]

  \matrix{
    \node[source] (hisparcbox) {$\rho:\mathfrak{sl_2} \rightarrow \mathfrak{g}$ ("Nahm data")};
      \& \node[datastore] (daq) {The set $S_N^+$}; \& \\

    \& \node[process] (buffer) {Irrep of $W[\mathfrak{g}](=W[\mathfrak{g}^L])$}; \& \\

      \& \node[datastore] (monitor) {$H^*(\mathcal{B}_{e'},\mathbb{C}$)};
      \& \node[sink] (datastore) {$\rho' :\mathfrak{sl_2} \rightarrow \mathfrak{g}^L$("Hitchin Data")}; \\
  };

  \draw[to] (hisparcbox) -- node[midway,above] {Null vector}
      node[midway,below] {conditions} (daq);
  \draw[to] (daq) -- node[midway,right] {MacDonald-Lusztig construction} (buffer);
  \draw[to] (buffer) --
      node[midway,right] {Springer's construction} (monitor);
  \draw[to] (monitor) -- node[midway,above] {}
      node[midway,below] {} (datastore);
\end{tikzpicture}
\caption{The general setup that is proposed for a physical interpretation of the map $d_{LS} : \rho  \rightarrow \rho'$. In this paper, only the case $\mathfrak{g} = \mathfrak{sl}_N$ is considered. Here, the Springer map is bijective and $\mathfrak{g} = \mathfrak{g}^L$ and hence $d_{LS}$ is an involution on the entire set of nilpotent orbits.}
\label{LSmap}
\end{figure}
\end{center}

\newpage
The interpretation of the map in cases where $\mathfrak{g}$ is not of type A is much more subtle. The anchor at the center of the figure \ref{LSmap} is no longer just a single irrep of $W[\mathfrak{g}]$. It could in general be a set of irreducible representations but with a unique \textit{special} representation in each set that occurs as the anchor. In the case of type A, all irreps are special and thus occur by themselves as anchors\footnote{A more precise way to say this is that in type A each representation has its own family/two-cell (as defined by Lusztig). In general, there may be more than one irrep associated to a family/two-cell. This more elaborate machinery is not needed in this paper. A follow up work will expand more on these themes \cite{Balasubramanian:2014}.}.

\begin{center}
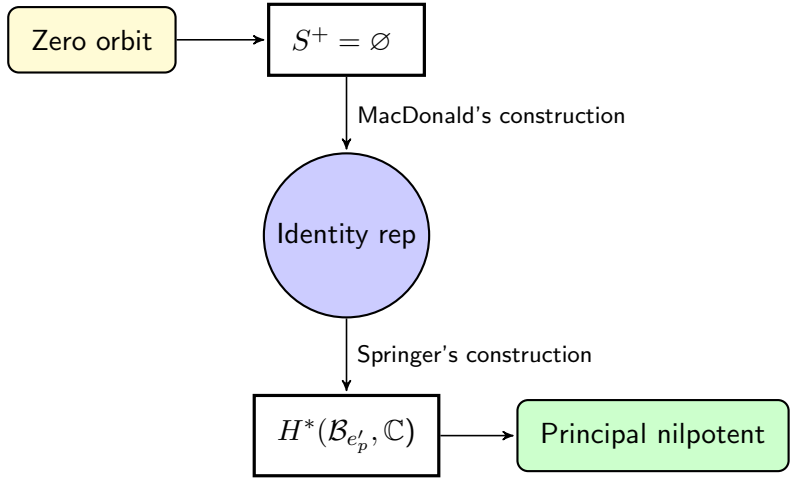
\begin{figure}[!h]
\begin{tikzpicture}[
  font=\sffamily,
  every matrix/.style={ampersand replacement=\&,column sep=1cm,row sep=1cm},
  source/.style={draw,thick,rounded corners,fill=yellow!20,inner sep=.3cm},
  process/.style={draw,thick,circle,fill=blue!20},
  sink/.style={source,fill=green!20},
  datastore/.style={draw,very thick,shape=rectangle,inner sep=.3cm},
  dots/.style={gray,scale=2},
  to/.style={->,>=stealth',shorten >=1pt,semithick,font=\sffamily\footnotesize},
  every node/.style={align=center}]

  \matrix{
    \node[source] (hisparcbox) {Zero orbit};
      \& \node[datastore] (daq) {$S^+={\varnothing}$ }; \& \\

    \& \node[process] (buffer) {Identity rep}; \& \\

      \& \node[datastore] (monitor) {$H^*(\mathcal{B}_{e'_p},\mathbb{C}$)};
      \& \node[sink] (datastore) {Principal nilpotent}; \\
  };

  \draw[to] (hisparcbox) -- node[midway,above] {}
      node[midway,below] {} (daq);
  \draw[to] (daq) -- node[midway,right] {MacDonald's construction} (buffer);
  \draw[to] (buffer) --
      node[midway,right] {Springer's construction} (monitor);
  \draw[to] (monitor) -- node[midway,above] {}
      node[midway,below] {} (datastore);
\end{tikzpicture}
\caption{A specific realization with the identity representation as anchor.}
\label{identity}
\end{figure}
\end{center}

\begin{center}
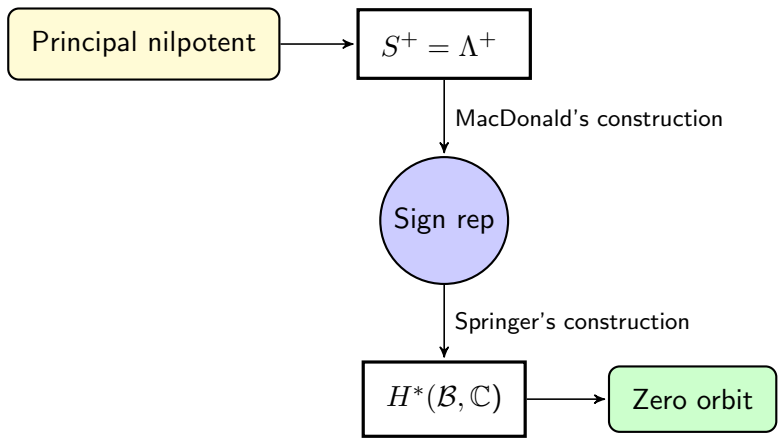
\begin{figure}[!h]
\begin{tikzpicture}[
  font=\sffamily,
  every matrix/.style={ampersand replacement=\&,column sep=1cm,row sep=1cm},
  source/.style={draw,thick,rounded corners,fill=yellow!20,inner sep=.3cm},
  process/.style={draw,thick,circle,fill=blue!20},
  sink/.style={source,fill=green!20},
  datastore/.style={draw,very thick,shape=rectangle,inner sep=.3cm},
  dots/.style={gray,scale=2},
  to/.style={->,>=stealth',shorten >=1pt,semithick,font=\sffamily\footnotesize},
  every node/.style={align=center}]

  \matrix{
    \node[source] (hisparcbox) {Principal nilpotent};
      \& \node[datastore] (daq) {$S^+=\Lambda^+$ }; \& \\

    \& \node[process] (buffer) {Sign rep}; \& \\

      \& \node[datastore] (monitor) {$H^*(\mathcal{B},\mathbb{C}$)};
      \& \node[sink] (datastore) {Zero orbit}; \\
  };

  \draw[to] (hisparcbox) -- node[midway,above] {}
      node[midway,below] {} (daq);
  \draw[to] (daq) -- node[midway,right] {MacDonald's construction} (buffer);
  \draw[to] (buffer) --
      node[midway,right] {Springer's construction} (monitor);
  \draw[to] (monitor) -- node[midway,above] {}
      node[midway,below] {} (datastore);
\end{tikzpicture}
\caption{A specific realization with the sign representation as anchor.}
\label{sign}
\end{figure}
\end{center}

Two of the simplest cases are given as evidence for such a setup in Figures \ref{identity} and \ref{sign}. As a more detailed example, the map for all the nilpotent orbits in $\mathfrak{sl}_4$ is recorded in Table \ref{sl4table}. For recording the irreps of $\mathcal{S}_4$, the standard partition notation is used. Each row in the table corresponds to a codimension two defect of the 6d theory. Also included are the values of some quantities that arise naturally when considering the local properties of these defects and the representation theory of Weyl groups.

\begin{center}
\begin{table}[!h]
\begin{center}
\begin{tabular}{|c|c|c|c|c|c|}
\hline 
 Nahm orbit  & $\mid \Lambda^+ \mid - \frac{dim(\mathcal{O}_N)}{2}$ & $r=Irr(W)$  & $\mid \Lambda^+ \mid - \frac{dim(\mathcal{O}_H)}{2}$ & $\frac{dim(\mathcal{O}_H)}{2}$ & Hitchin orbit   \\ 
 ($\mathcal{O}_N$) & = $(n_h - n_v)$ & & =$a(r)$ & =$d$ & ($\mathcal{O}_H$)  \\
\hline 
 $[1^4]$ & 6 & $[4]$ & 0 & 6 &$[4]$ \\ 
\hline 
  $[2 ,1^2]$ & 3 & $[3,1]$ & 1 & 5 &$[3,1]$ \\ 
\hline 
  $[2,2]$ & 2 & $[2,2]$ & 2 & 4 & $[2,2]$ \\ 
\hline 
 $[3,1]$ & 1 & $[2,1^2]$ & 3 & 3 &$[2, 1^2]$ \\ 
\hline 
$[4]$ & 0 & $[1^4]$ & 6 & 0 & $[1^4]$ \\ 
\hline 
\end{tabular} 
\label{sl4table}
\end{center}
\caption{A table encoding the Lusztig-Spaltenstein map for all the nilpotent orbits for $\mathfrak{g} = \mathfrak{g}^L = \mathfrak{sl}_4$. The labels for the Nahm and Hitchin data are the partition types associated to the corresponding nilpotent orbits. \textit{Notation} : $a(r)$ is the $a$-invariant attached to an irreducible representation of the Weyl group by Lusztig, $d$ is the total Coulomb branch dimension and $n_h, n_v$ are the contribution from each defect to the effective numbers of the hypermultiplets and vector multiplets.}
\end{table}
\end{center}

Note that the quantities $n_h - n_v$ and $a(r)$ also have a direct interpretation in Springer theory,
\begin{eqnarray}
n_h - n_v = \text{dim}_{\mathbb{C}} (\mathcal{B}_N), \\
a(r) = \text{dim}_{\mathbb{C}} (\mathcal{B}_H),
\end{eqnarray}
where $\mathcal{B}_N$ and $\mathcal{B}_H$ are Springer fibers associated to the Nahm orbit (given by $\rho$ in $\mathfrak{g}$) and Hitchin orbit (given by $\rho'$ in $\mathfrak{g}^L$)respectively. In the language of Hitchin systems, these fibers would correspond to the Hitchin fiber (or the dual Hitchin fiber) above a ramification point on the base where the Higgs field $\phi$ has a simple pole with a nilpotent residue that belongs in $\mathcal{O}_N$ (or $\mathcal{O}_H$).

See \cite{barbasch1985unipotent} for a similar scenario where Jacobson-Morozov theory and the Springer theory are used on $\mathfrak{g}$ and $\mathfrak{g}^*$ respectively. The relationship between geometric approaches to Springer theory (and its generalizations) and Hitchin systems have been explored recently in the context of the geometric Langlands program \cite{Gukov:2006jk,nadler2011springer,Ben-Zvi:2013aka}.

\subsection{Examples of free theories : $A_2$ tinkertoys}

The overall scale factor calculation from a Toda perspective is much simplified when the corresponding 4d theory is just a free theory of hypermultiplets. These are the theories for which the total $n_v$ is zero. Recall that this quantity is defined as 
\begin{equation}
n_v = \sum_i n_v^{i} + n_v^{\text{global}},
\end{equation}
where $n_v^{global}$ is defined as
\begin{equation}
n_v^{global} = (1-g) (\frac{4}{3} \hat{h} \text{dim} (G)  + \text{rank} (G)),
\end{equation}
and $n_v^i$ is given by \ref{nvlocal}.
In the tinkertoy terminology, these are called free fixtures \cite{Chacaltana:2010ks}. Let us consider one of the free fixtures that occur in the $A_2$ theory and understand how the $n_h$ contribution to the scale factor is encoded in the corresponding Toda correlator. Specializing the Toda action on a disc to this case,
\begin{equation}
S_{T,\text{disc}} =  \frac{1}{72 \pi} \int_D \sqrt{\hat{g}} d^2 z \bigg ( \frac{1}{2} \hat{g}^{ab} \partial_a \phi \partial_b \phi   + \sum_{i=1}^{i=2} 2 \pi \Lambda e^{2b(e_i,\phi)}  \bigg) + \frac{1}{6\pi}\int_{\partial D} (Q,\phi) d \theta + \frac{2}{3} (Q,Q) \log (R/R_0).
\label{actionToda}
\end{equation}
There are two regular punctures to consider when dealing with the $A_2$ family of theories of class $S$. The root space is two dimensional and is spanned by the simple roots $\vec{e_1}, \vec{e_2}$. The roots are normalized so that the the entries in scalar product matrix $K_{i,j} = (\vec{e_i},\vec{e_j})$ are given by $K_{ii} = 2, K_{12} = K_{21} = -1$. The set of positive roots is  $ \vec{e}>0=\{\vec{e}_1, \vec{e}_2, \vec{e}_3\}$ where $\vec{e_3} = \vec{e_1} + \vec{e_2}$. The fundamental weights are $\vec{\omega}_1, \vec{\omega}_2$ and they obey $(\vec{\vec{\omega}}_i,\vec{e}_j)=\delta_{ij}$. As usual, $\vec{\rho}$ is half the sum of positive roots and $h_i$ (the weights of the fundamental representation) are given by
\begin{eqnarray}
 h_1 &=& \vec{\omega}_1,\\
h_2 &=& h_1 - e_1,\\
h_3 &=& h_2 - e_2. 
\end{eqnarray}
The maximal puncture corresponds to a Toda primary $\mathcal{O}^{max} _{\vec{p}} = \exp{(\vec{p}}.\vec{\phi)}$ where $\vec{p}$ is valued in the dual of the lie algebra. Writing $\vec{p} = \alpha_1 \vec{\omega_1} + \alpha_2\vec{\omega_2}$, it is seen that a general Toda primary has two complex numbers as parameters. In the $A_2$ Toda case, there is yet another puncture which corresponds to $\mathcal{O}^{min}_{\vec{p}} = \exp (\vec{p}.\vec{\phi})$ where $\vec{p}$ in constrained to $\vec{p} = \chi \vec{\omega}_2$ (or equivalently $\chi \omega_1$).


\subsubsection{$V[\mathfrak{sl_3}]_{0,([2,1],[1^3],[1^3])}$}
\begin{figure}
\begin{center}
\includegraphics{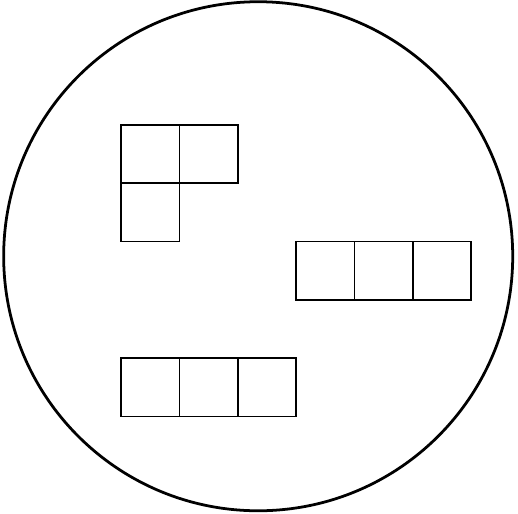}
\caption{$A_2$ theory on a sphere with one minimal and two maximal punctures }
\end{center}
\end{figure}
The three point function with one argument taking a semi-degenerate value was obtained in \cite{Fateev:2007ab}. It is given by
\begin{equation*}
 V[\mathfrak{sl_3}]_{0,([2,1],[1^3],[1^3])} = C(\alpha_1,\alpha_2,\alpha_3) |z_{12}|^{-2(\Delta_1 + \Delta_2 -\Delta_3)} |z_{13}|^{-2(\Delta_1 + \Delta_3 - \Delta_2)} |z_{23}|^{-2(\Delta_2 + \Delta_3 -\Delta_1)},
\end{equation*}
where
\begin{eqnarray*}
 C (\chi \vec{\omega_2},\vec{p_1},\vec{p_2}) &=& \bigg[\pi \Lambda  \gamma(b^2) b^{2-2b^2} \bigg]^{(Q-\sum_i \alpha_i)/b} \times \\ && \frac{\Upsilon(b)^{n-1}\Upsilon(\chi)\prod_{\vec{e} > 0} \Upsilon((\vec{Q}-\vec{p_1}).\vec{e})\Upsilon((\vec{Q}-\vec{p_2}).\vec{e})}{\prod_{i=1,j=1}^{i=3,j=3} \Upsilon \bigg( \frac{\rho}{2} + (\vec{p_1}- \vec{Q}).\vec{h_i} + (\vec{p_2}- \vec{Q}).\vec{h_j} \bigg)}.
\end{eqnarray*}
As was the case with the three punctured sphere in the Liouville case, the poles comes from the $\Upsilon$ functions in the denominator and these correspond to the screening conditions. For the $A_2$ case, there are two primitive screening conditions 
\begin{eqnarray}
( \rho \vec{\omega}_2 + \vec{p_2} + \vec{p_3}).\vec{\omega}_1 = \Omega_{m,n}, \\
( \rho \vec{\omega}_2 + \vec{p_2} + \vec{p_3}).\vec{\omega}_2 = \Omega_{m.n},
\end{eqnarray}
and the rest are obtained by applying the two Weyl relations and identifying screening conditions that differ only by an overall Weyl reflection. The two reflections act by
\begin{eqnarray}
 \sigma_1 : \vec{p} \rightarrow ((2 \vec{Q} - \vec{p}).\vec{e_1} ) \vec{e_1}, \\
\sigma_2 :  \vec{p} \rightarrow ((2 \vec{Q} - \vec{p}).\vec{e_2} ) \vec{e_2}.
\end{eqnarray}
where $\vec{Q} = Q \vec{\rho}$ and $Q = b + b^{-1}$ as before.
The number of distinct screening conditions agrees with the assignment $n_h = 9$ for this fixture.  As with Liouville correlators, we define a stripped version,
\begin{equation}
 \hat{V}[\mathfrak{sl_3}]_{0,([2,1],[1^3],[1^3])} =  \frac{V[\mathfrak{sl_3}]_{0,([2,1],[1^3],[1^3])}}{\Upsilon(b)^{n-1}\Upsilon(\chi)\prod_{\vec{e} > 0} \Upsilon((\vec{Q}-\vec{p_1}).\vec{e})\Upsilon((\vec{Q}-\vec{p_2}).\vec{e})}.
\end{equation}

The scale factor for the stripped correlator comes from combining the anomalous scaling of the nine $\Upsilon$ functions that enforce the screening conditions. This gives,
\begin{equation}
\hat{V}[\mathfrak{sl_3}]_{0,([2,1],[1^3],[1^3])} = \mu^{9/6}\hat{V}[\mathfrak{sl_3}]_{0,([2,1],[1^3],[1^3])}^{R_0=1}
\end{equation}

The argument can also be inverted in the sense that the knowledge of the scale factor for the stripped correlator corresponding to a free theory can be used to predict the analytical structure (=number of polar divisors) of the corresponding Toda three point function. Two such families are discussed below as examples. It is worth emphasizing that this is by no means an exhaustive list.

\subsection{Families of free fixtures and corresponding Toda correlators}
\label{freefamilies}

In the literature on Toda theories, the only correlation functions for which the analytical structure is explicitly known is the Fateev-Litvinov family \cite{Fateev:2007ab}. These correspond to the family of free fixtures that will be called $f_N$. They correspond to $N^2$ free hypermultiplets transforming in the $(N,\bar{N})$ representation of the flavor symmetry group. This data is reflected in the fact that the FL family of Toda correlators have $N^2$ polar divisors with the exact same representation structure. That this should be the case could have been inferred from knowing the scale factor assigned to this correlator and deducing the value of $n_h$ from that. Recall that for a free theory, $n_h = 24a$. The conjecture is that $n_h$ is the number of polar divisors for the corresponding Toda correlator.
For the Toda correlators corresponding to other families of free fixtures, corresponding results do not seem to be available in the literature. But, knowing the corresponding scale factor values along with the representation data \citep{Chacaltana:2010ks} , the analytical form of these correlators can be conjectured. This can be done for $\text{any}$ family of free fixtures using the following formula
\begin{equation}
n_h = \sum_i n_h^{i} - 16 (\rho,\rho),
\end{equation}
where $n_h^{i}$ is the contribution from each primary insertion and can be deduced from the scale factor in (\ref{primarymap2}). The last term is the global contribution from the sphere with $\rho$ denoting the Weyl vector. Let us now look at a couple of examples to understand what is meant by families of free theories.

\subsubsection{$f_n$}
\begin{figure}[!h]
\begin{center}
\includegraphics{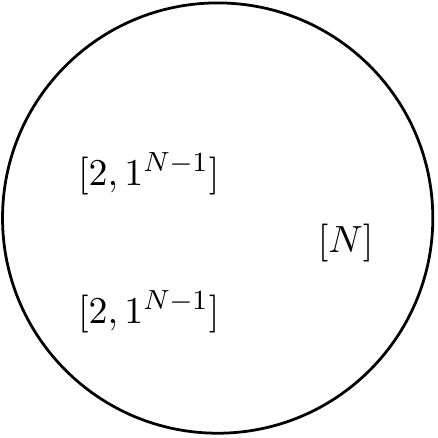}
\caption{The $f_N$ family of free fixtures corresponding to the Fateev-Litvinov family of Toda three point functions.}
\end{center}
\end{figure}
This is the Fateev-Litvinov family corresponding to $N^2$ polar divisors. This does correspond to the $n_h$ value associated to this free fixture. In the uniform notation used for Toda correlators, this would correspond to $V[\mathfrak{sl}_N]_{0,([2,1^{N-1}],[2,1^{N-1}],[N])}$.
\subsubsection{$g_n$}
This is a new family $V[\mathfrak{sl}_N]_{0,([2^2,1^{N-2}],[3,2,1^{N-2}],[N])}$ of three point functions for which the analytical structure can be conjectured based on the Tinkertoy constructions. This family has $n_h = \frac{1}{6}N^3 - \frac{3}{2}N^2 + \frac{28}{3}N - 10$ and this number should equal the number of polar divisors (built out of $\Upsilon$ functions as in the case of $f_N$). From a purely Toda perspective, requiring that the poles arise only from the screening conditions (and its Weyl reflections) for this correlator should lead to the same result. 
\begin{figure}[!h]
\begin{center}
\includegraphics{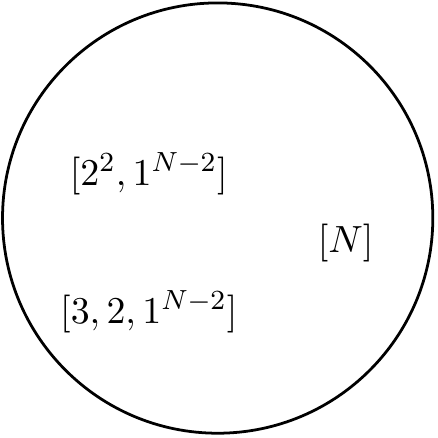}
\caption{The $g_N$ family of free fixtures corresponding to a family of Toda three point functions.}
\end{center}
\end{figure}
\newpage
\section{Scale factors in Toda correlators II : Interacting theories}
\label{Antheories2}

\subsection{Factorization in Toda theories}

Apart from observing that $Z_{\mathbb{S}^4}$ matches with the Liouville correlators, AGT also noted that the identities satisfied by CFT correlators with single $T(z)$ insertions can be understood as a deformed version of the Seiberg-Witten curve. For example, $\mathbf{T}(z)$ insertions in Liouville correlators on the sphere obey the following identity,
\begin{equation*}
 \langle \mathbf{T} (z) \prod_i \mathcal{O}_i (z_i) \rangle = \sum_i \bigg( \frac{\Delta_i}{(z-z_i)^2}+ \frac{\mathcal{L}_{-1}}{z-z_i}\ \bigg) \langle \prod_i \mathcal{O}_i (z_i) \rangle.
\end{equation*}
These are what are called the conformal Ward identities. An immediate consequence of this is that correlation functions of descendants (defined to be states obtained by acting on $\mathcal{O}_i$ by modes of $T(z)$ or $\bar{T}(\bar{z})$) are fully determined in terms of the correlation functions of the primaries. 

Let us now define the following quadratic differential,
\begin{equation*}
 \phi_2 (z)dz^2 = - \frac{\langle \mathbf{T}(z) \prod_i \mathcal{O}_i (z_i) \rangle }{\langle \prod_i \mathcal{O}_i (z_i) \rangle}.
\end{equation*}
In a suitable  limit, the conjecture \cite{Alday:2009aq} is that
\begin{equation*}
 \phi_2 (z)dz^2  \rightarrow \phi_2 ^{SW}.
\end{equation*}
In the general Toda case, the full chiral algebra has more such identities that arise from insertions of the higher spin tensors $\mathbf{W}^n(z)$, $n>2$. However, the $\mathcal{W}$-Ward identities fail to  determine the correlation functions with descendants completely in terms of the correlators of primaries. One can define a number that quantifies the nature of this failure. This number turns out to be related to the total Coulomb branch dimension. As an example, consider the three point in $A_2$ Toda theory together with all its descendants. 
\begin{equation*}
 D(V_{0,\{0,3\}}) = \langle \prod_{i=1}^{3} D_i \mathcal{O}_{\vec{p}(z_i)}\rangle,
\end{equation*}
where $D_i$ is a product of the modes of the operators $T(z)$ and $W_3(z)$. The primaries obey
\begin{eqnarray*}
 \mathbf{T}(z) \mathcal{O}(w) &=& \frac{\Delta \mathcal{O}(w)}{(z-w)^2} + \frac{\partial \mathcal{O} (w)}{(z-w)} + \text{non-singular}\\
\mathbf{W}^3(z) \mathcal{O} (w) &=& \frac{\Delta^{(3)} \mathcal{O}(w)}{(z-w)^3} + \frac{W^{(3)}_{-1}\mathcal{O}(w)}{(z-w)^2} + \frac{W^{(3)}_{-2} \mathcal{O}(w)}{(z-w)} + \text{non-singular}.
\end{eqnarray*}
Observe that $D(V_{0,\{0,3\}})$ obeys a set of local ward identities. These can be  obtained by inserting $\int_\infty f_k W_k (z) =0$ into the correlator where $f_s$ is a meromorphic function with poles at $z=z_i$. Using the local ward identities, all correlators in the family can be written in terms of those of the form $D_0(V_{0,\{0,3\}})$ where $D_0 = \{ L_{-1}, W_{-1}, W_{-2} \}$. The total number of linearly independent correlators in the set $D_0(V_{0,\{0,3\}})$ is nine (three $D_0$s for each primary). Imposing the global ward identities further constrains this set of correlators. The total number of global ward identities is 8 in the $W_3$ case. This shows that $W$-symmetry fails to determine the correlators of descendants completely in terms of that of the primaries. A representative of the set of correlation functions than cannot be linearly related to $V_{0,\{,0,3\}}$ is $\langle W_{-1}^k \mathcal{O}_1
\mathcal{O}_2 \mathcal{O}_3 \rangle$. Let us assign Coulomb branch dimension as $d = 9- 8=1$ to this family. It is easy to see that when one of the primaries is semi-degenerate, the total Coulomb branch dimension is zero. This is because the null vector takes the following form
\begin{equation*}
 (L_{-1} - \frac{3}{2} W_{-1} ) | \mathcal{O}_1 \rangle = 0.
\end{equation*}
This can be used to turn the $W_{-1}$ to a $L_{-1}$. So, the family $D_0(V_{0,\{1,2\}})$ actually has no Coulomb branch (Coulomb branch dimension is zero). Using the spectrum of semi-degenerate operators in Toda theory and null vector conditions that they obey, this dimension can be calculated for any such family. This matches the corresponding 4d field theory's Coulomb branch dimension. For a similar count of equations, see \cite{Fateev:2010za} and \cite{Kozcaz:2010af}.
One can also define a finer quantity, namely the graded Coulomb branch $d_k$. This is related to the quantity called $n_v$ by the following formula
\begin{equation}
n_v = \sum_k (2k-1) d_k.
\end{equation}
Recall here the definition of $n_v^{global}$,
\begin{equation}
n_v^{global} = (1-g) (\frac{4}{3} \hat{h} \text{dim} (G)  + \text{rank} (G)),
\end{equation}
where $\hat{h}$ is the dual Coxeter number and $G = SU(N)$ for all cases considered here.
Some practice with the appearance of smaller gauge groups in the various limits of the corresponding 4d theories leads us to propose the following criteria for a full factorization in Toda theory. This corresponds to the appearance of an $SU(N)$ gauge group in the four dimensional theory. Take the degeneration limit where punctures $\alpha_i$ appear one side of the channel and punctures $\beta_j$ appear on the other side. Construct the following quantities,
\begin{eqnarray}
X_{\alpha}  \equiv\sum_i n_v^{\alpha_i} + n_v^{max} + n_v ^{global,g=0}, \\ 
X_{\beta} \equiv \sum_i n_v^{\beta_i} + n_v^{max} + n_v ^{global,g=0}.
\end{eqnarray}
If and only if $X_\alpha,X_\beta \geq 0$, there is full factorization for the Toda\footnote{For the case of Liouville, this reduces to the familiar condition for a macroscopic state to occur in the factorization channel\cite{Polchinski:1990mh,Seiberg:1990eb}.} correlator. Exactly which subgroup appears as the gauge group in a channel where one of the quantities $X_\alpha,X_\beta$ become negative requires more detailed analysis involving the exact Toda correlators. This seems possible to carry out only in a limited number of cases (see example below). On the four dimensional side, this data has been determined in \cite{Chacaltana:2010ks} using constraints that come from requiring Coulomb branch diagnostics like the graded $d_k$ to match in all factorization limits.  A physical interpretation of this phenomenon using the properties of the Higgs branch has been given in \cite{Gaiotto:2011xs}. 
\subsubsection{A conjecture}
With the experience of examples worked out so far and based on the general physical expectation that the Euler anomaly should be encoded as a scale factor in the sphere partition function of any conformal class $\mathcal{S}$ theory, one can formulate the following conjecture.
\label{conjecture}
\begin{conjecture}
 Let $\hat{V}[\mathfrak{g}]_{g,(\{\mathcal{O}_N^i\})}$ be the stripped Toda correlator corresponding to the sphere partition function of class $\mathcal{S}$ SCFT (with mass deformation parameters $m_i$) obtained by taking theory $\mathscr{X}[\mathfrak{g}]$ on Riemann surface of genus $g$ with $n$ punctures along with $n$ codimension two defects (with Nahm labels $\{\mathcal{O}_N^i \},i=1 \ldots n$) placed at the punctures. Let the Euler anomaly of the SCFT be $a$ and the inverse radius of the four sphere on which the SCFT is formulated be $\mu$.
 Then, $\hat{V}[\mathfrak{g}]_{g,(\{\mathcal{O}_N^i\})}$=$\mu^{4a}(\hat{V}[\mathfrak{g}]_{g,(\{\mathcal{O}_N^i)\}})_{R_0=1}$ in the $m_i \rightarrow 0$ limit, irrespective of the factorization limit in which the scale factor is calculated.
\end{conjecture}

The stripped correlator $\hat{V}$ in the general case is defined to be
\begin{equation}
\hat{V}[\mathfrak{g}]_{g,(\{\mathcal{O}_N^i\})} = \frac{ V[\mathfrak{g}]_{g,(\{\mathcal{O}_N^i\})} \Upsilon(b)^{rank(\mathfrak{g})(g-1)} }{\prod_i D^0_i},
\label{stripped}
\end{equation}
where $\prod_i D^0_i$ is the collection of all factors in the correlator $V[\mathfrak{g}]_{g,(\{\mathcal{O}_N^i\})}$ that become identically zero in the $m_i \rightarrow 0$ limit. In certain familiar cases, the factors $D^0_i$ have an expression in terms of $\Upsilon$ functions. In the more general cases, the inverse of the stripped correlator may be best viewed as an iterated residue \footnote{$IRes(\ldots)_{m_i\rightarrow0}=Res(Res(\ldots)_{m_1\rightarrow 0})_{m_2 \rightarrow 0}$ and so on.},
\begin{equation}
\hat{V}[\mathfrak{g}]_{g,(\{\mathcal{O}_N^i\})}^{-1} =\frac{ \text{IRes}\bigg(V[\mathfrak{g}]_{g,(\{\mathcal{O}_N^i\})}^{-1}\bigg)_{m_i \rightarrow 0}}{\Upsilon(b)^{rank(\mathfrak{g})(g-1)} }.
\end{equation}

Following the intuition from the path integral argument for the three point function in the Liouville case, one expects that the parameter $\mu$ can be understood to be the dimensionful parameter that enters in the definition of the regularized correlator.
When the correlator is such that every factorization limit involves a channel with $X_\alpha,X_\beta \geq 0$, it is immediate that the scale factor is independent of the limit in which it is evaluated. When this is not the case, the above statement is a non-trivial constraint on the nature of the state appearing in the factorization channel (For such an example, see Section \ref{asymmetric} below). The above conjecture is stated for arbitrary $\mathfrak{g}$ since it is expected to hold in all the cases. This paper provides a concrete setup for the case $\mathfrak{g}=A_n$. 
\subsection{Examples : Theories with a known Lagrangian}

\subsubsection{$V[\mathfrak{sl_3}]_{0,([2,1],[2,1],[1^3],[1^3])}$ in its symmetric limit}
\begin{figure}
\begin{center}
\includegraphics{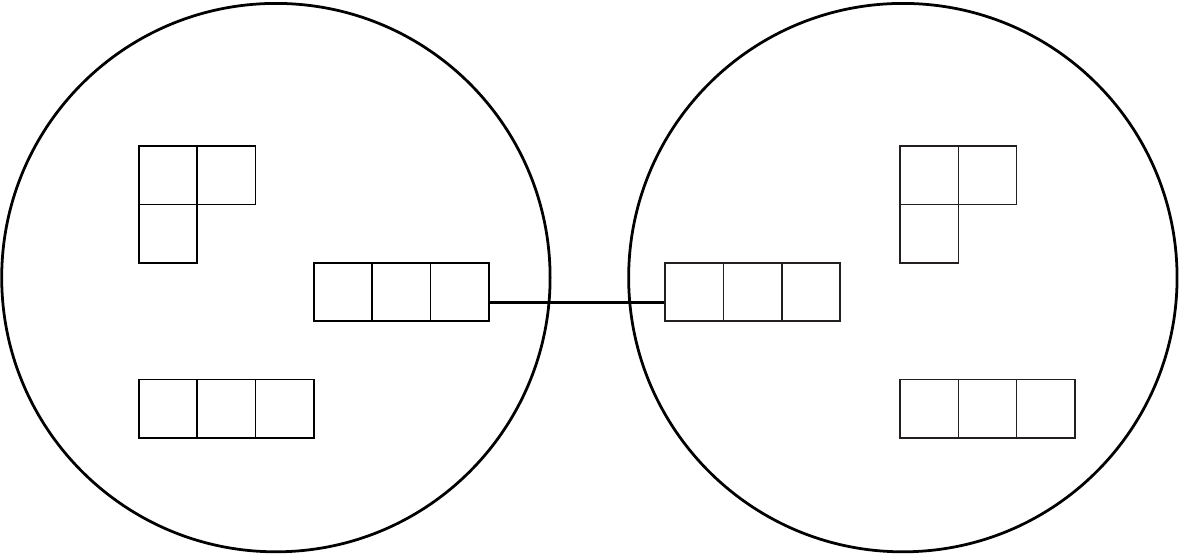}

\caption{The $A_2$ theory on a sphere with two minimal and two maximal in the symmetric limit. }
\end{center}
\end{figure}

Since the most general three point function is not known in closed form, this four point function is written in the factoring limit that allows us to express it in terms of the $f_N$ family of three point functions in the following way
\begin{eqnarray*}
 V[\mathfrak{sl_3}]_{0,([2,1],[2,1],[1^3],[1^3])} (\rho \vec{\omega_2},\sigma \vec{\omega_2},\vec{p_1},\vec{p_2}) &=&  \bigint_{\vec{p}\in \vec{Q} + i(s_1^+\vec{\omega_1}+ s_2^+\vec{\omega_2})} d \vec{p} C (\rho \vec{\omega_2},\vec{p_1},\vec{p}) C (\vec{Q}-\vec{p},\vec{p_2},\sigma \vec{\omega_2}) \\ && \mathcal{F}_{\mathfrak{sl_3}} \bigg[\begin{array}{ll}
\vec{p_1} & \vec{p_2}\\
\chi \vec{\omega_2} & \sigma \vec{\omega_2}
                                                                                                                                                                                                                 \end{array} \bigg] (\vec{\alpha},z_i)\mathcal{F}_{\mathfrak{sl_3}} \bigg[\begin{array}{ll}
\vec{p_1} & \vec{p_2}\\
\chi \vec{\omega_2} & \sigma \vec{\omega_2}
                                                                                                                                                                                                                 \end{array} \bigg] (\vec{Q}-\vec{\alpha},\bar{z_i}),
\end{eqnarray*}
where the three point function belong to the Fateev-Litvinov family $f_N$.
The dependence of the conformal blocks on the momenta is through the dimensions $\Delta_{\vec{p}}, \Delta^{(3)}_{\vec{p}}$. These are given by
\begin{eqnarray}
\Delta_{\vec{p}} &= &\frac{(2 \vec{Q} - \vec{p}).\vec{p}}{2}, \\
\Delta^{(3)}_{\vec{p}} &=& i \sqrt{\frac{48}{22+5c}} (\vec{p} - \vec{Q},h_1) (\vec{p}-\vec{Q},h_2)(\vec{p} - \vec{Q},h_3)  .
\end{eqnarray}
Proceeding as in the case of the four point function for Liouville, one can rewrite $\Upsilon$ functions in the numerator in terms of the $H$ functions making the Vandermonde explicit. This gives an integration of the form $\int da_1 d a_2 (a_1^2 + a_2^2)(a_1^4 + a_2^4 )$ implying $n_v = 8$ (as expected for a gauge theory with gauge group $SU(3)$). Defining $\hat{V}[\mathfrak{sl_3}]_{0,([2,1],[2,1],[1^3],[1^3])}$ as in \ref{stripped} and collecting the anomalous scaling factors,
\begin{equation}
\hat{V}[\mathfrak{sl_3}]_{0,([2,1],[2,1],[1^3],[1^3])}  = \mu^{29/3}(\hat{V}[\mathfrak{sl_3}]_{0,([2,1],[2,1],[1^3],[1^3])} )_{R_0=1}.
\end{equation}
The value of $4a$ is correctly reproduced.

\subsubsection{$V[\mathfrak{sl_3}]_{1,([2,1])}$}
\begin{figure}
\begin{center}
\includegraphics{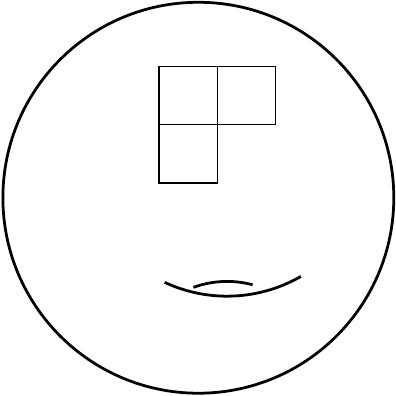}
\caption{$A_2$ theory on a torus with one minimal puncture }
\end{center}
\end{figure}
This is the correlator that pertains $Z_{\mathbb{S}^4}$ of $SU(3)$ gauge group with an adjoint hypermultiplet and a free hyper. It has the following expression,
\begin{equation*}
V[\mathfrak{sl_3}]_{1,([2,1])}(\chi \vec{\omega_2})  = \bigint d \vec{p} \frac{\Upsilon(b)^{n-1}\Upsilon(\rho)\prod_{\vec{e} > 0} \Upsilon((\vec{Q}-\vec{p}).\vec{e})\Upsilon((\vec{Q}+\vec{p}).\vec{e})}{\prod_{i=1,j=1}^{i=3,j=3} \Upsilon \bigg( \frac{\rho}{2} + (\vec{p}- \vec{Q}).\vec{h_i} + (\vec{Q}-\vec{p}).\vec{h_j} \bigg)} \mathcal{F}_{\mathfrak{sl_3}}^{g=1} [\chi \omega_2 , \vec{p} ] .
\end{equation*}
Again, defining $\hat{V}[\mathfrak{sl_3}]_{1,([2,1])}$ following \ref{stripped} and collecting anomalous scale factors,
\begin{equation}
\hat{V}[\mathfrak{sl_3}]_{1,([2,1])} = \mu^{49/6} (\hat{V}[\mathfrak{sl_3}]_{1,([2,1])})_{R_0=1}.
\end{equation}
Ignoring the contribution from the decoupled abelian vector multiplets reproduces the expected value for $4a$.

\subsection{Examples : Theories with no known Lagrangian description}

\subsection{$V[\mathfrak{sl_3}]_{0,([2,1],[2,1],[1^3],[1^3])}$ in its asymmetric limit}
\label{asymmetric}

Let us now consider this correlator in the limit where two minimal punctures are on one side and the two maximal punctures are on the other side of the factorization channel. The duality between the corresponding four dimensional theories (that arise in the two limits) was discovered by Argyres-Seiberg \cite{Argyres:2007cn}.

In this limit, $X_\alpha <0, X_\beta >0$. So, the condition for a full factorization is not satisfied. In its other limit, we have already seen that this theory has $n_h = 18,n_v=8$ (with the corresponding implications for the three point functions appearing in the symmetric limit). To understand the asymmetric limit, let us write the four point function in the following form

\begin{eqnarray*}
V[\mathfrak{sl_3}]_{0,([2,1],[2,1],[1^3],[1^3])} (\rho \vec{\omega_2},\sigma \vec{\omega_2},\vec{p_1},\vec{p_2}) &=&  \bigint_{\vec{p}\in \vec{Q} + i(s_1^+\vec{\omega_1}+ s_2^+\vec{\omega_2})} d \vec{p} C (\rho \vec{\omega_2},\sigma \vec{\omega_2},\vec{p}) C (\vec{Q}-\vec{p},\vec{p_2},\vec{p_1}) \\ && \mathcal W_2 \bigg[\begin{array}{ll}
\vec{p_1} & \vec{p_2}\\
\chi \vec{\omega_2} & \sigma \vec{\omega_2}
                                                                                                                                                                                                                 \end{array} \bigg] (\vec{\alpha},z_i)\mathcal W_2 \bigg[\begin{array}{ll}
\vec{p_1} & \vec{p_2}\\
\chi \vec{\omega_2} & \sigma \vec{\omega_2}
                                                                                                                                                                                                                 \end{array} \bigg] (\vec{Q}-\vec{\alpha},\bar{z_i}).
\end{eqnarray*}

\begin{figure}
\begin{center}
\includegraphics{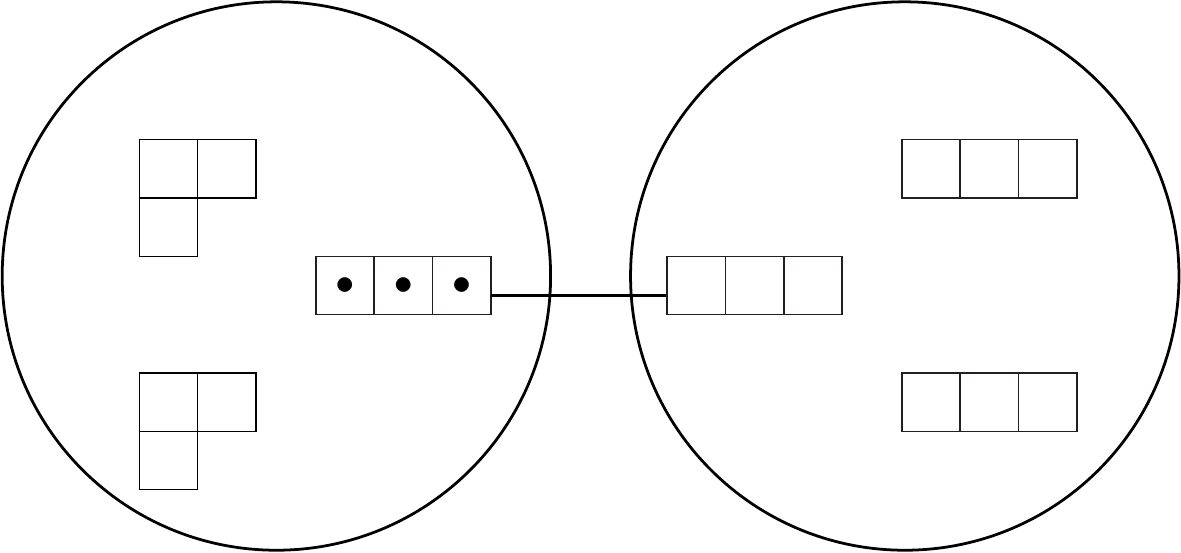}
\caption{The $A_2$ theory on a sphere with two minimal and two maximal in the asymmetric limit. }
\end{center}
\end{figure}

Here, the three point function $C (\rho \vec{\omega_2},\sigma \vec{\omega_2},\vec{p})$ can be understood as a limit of the Fateev Litvinov family $f_N$ where one of the maximal punctures is made minimal. When this is done, the three point function becomes identically zero except when the following condition is obeyed \cite{Kanno:2010kj,Drukker:2010vg},
\begin{equation}
w  - w_1 + w_2 + \frac{3}{2} \bigg(\frac{w_1}{\Delta_1} - \frac{w_2}{\Delta_2}\bigg) (\Delta - \Delta_1 - \Delta_2)  = 0.
\end{equation}
In the above equation the cubic invariant is referred to as $w$ instead of $\Delta^{(3)}$ to avoid confusion with the subscripts. The above condition restricts the channel momentum to a one dimensional subspace of the most general macroscopic Toda state. This corresponds to the choice of a $SU(2)$ subgroup. After canceling factors between the numerator and the denominator of the $f_N$ correlator (specialized to $N=3$) and using the properties of the $\Upsilon$ functions, the measure for the channel integral is seen to be of the form $a^2 da$. One would like to account for the scale factor in this limit. The $n_h$ contribution is easy to account for since this arises only from the local contributions of the punctures and the global contribution of the sphere. $n_v$ on the other hand is non-trivial. From the factorization channel, we get $n_v=3$ (as opposed to $n_v = 8$ from the factorization channel in the symmetric limit). This implies that the stripped three point function corresponding to three maximal punctures has a scale factor that corresponds to $n_h=16, n_v = 5$. 

This discussion aims to be nothing more than a poor substitute for an analysis of the factorization problem in Toda theories. It was included to provide an example of how the accounting for the scale factor could be different in the various factorization limits. It is examples like this that make the conjecture in Section \ref{conjecture} a non-trivial constraint on Toda factorization.

\subsection{$V[\mathfrak{sl_3}]_{0,([1^3],[1^3],[1^3])}$}

\begin{figure}
\begin{center}
\includegraphics{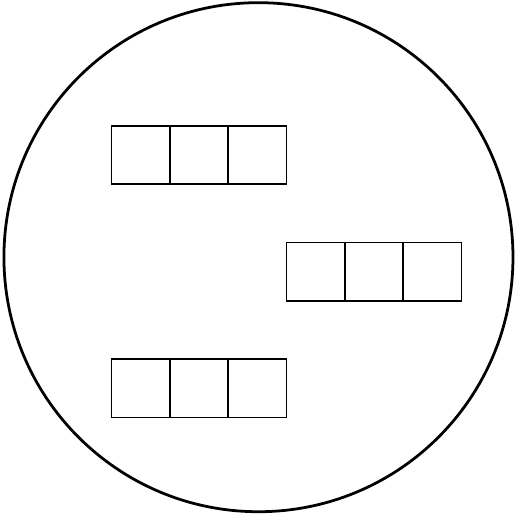}
\caption{$A_2$ theory on a sphere with three maximal punctures }
\end{center}
\label{T3}
\end{figure}
Not much is known in closed form for this correlator (Fig \ref{T3}). Integral expressions for this correlator are available under some special limits. See \cite{Fateev:2007ab,Fateev:2008bm} for the state of the art on Toda computations.  Note that this is the correlator corresponding to the partition function on $\mathbb{S}^4$ of the $\mathcal{T}_3$ theory. This correlator arises in a 'decoupling limit' of the previous example where two minimal punctures are collided and replaced with a maximal puncture. As discussed, the scale factor for the stripped correlator in this case should correspond to $n_h=16, n_v=5$.
\section{Summary}

It is argued in this paper that the Euler anomaly of a 4d SCFT belonging to class $\mathcal{S}$ is encoded in the scale factors of the corresponding stripped Liouville/Toda correlators. This factor is always of the form $\mu^{4a}$ where $a$ is the Euler anomaly and the quantity $\mu$ can be identified with the inverse radius of the four sphere on which the theory is formulated.  The quantity $a$ has a parameterization in terms of quantities $n_h,n_v$ (given in \ref{eulercharge}). The parameterization of $a$ by $n_h$ and $n_v$ is convenient since the two types of contributions to $a$ arise differently in the Liouville/Toda context\footnote{This is obviously so in the 4d theories with Lagrangian description. So, it is perhaps not a surprising feature.}, 
\begin{itemize}
\item The local $n_h$ contribution arises from the scale factors in the relationship between the Toda and WZW primaries while the global $n_h$ factor arises from the boundary term associated to the curvature insertion in the Toda action on the disc,
\item The $n_v$ contributions arise from every factorization channel (when there is one) and from the 'strongly coupled' SCFTs. The contribution from the former is straightforward to pin down while the latter is known by requiring consistency with crossing symmetries (S-dualities in the four dimensional context).
\end{itemize}

 The above setup should be contrasted with how these quantities are calculated in the four dimensional context in (\ref{NhandNv}). Requiring that they agree is then a non-trivial constraint on Toda factorization and a conjecture was outlined to this effect in Section \ref{conjecture}. When the total $n_v$ contribution is zero, the corresponding four dimensional theory is a free theory with $n_h$ hypermultiplets. The relationship between the scale factor in such theories and the analytical structure of the Toda correlator allows one to make predictions for the number of polar divisors in certain Toda correlators. Some examples of this were outlined in Section \ref{freefamilies}.

As briefly alluded to in the Introduction, the class of theories studied here have attracted attention from various different points of view. It is natural to consider the connections of those with the setup of this paper.  The conjecture that is provided for the scale factor should follow automatically if crossing symmetry for Toda theories is proved. In the case of Liouville CFT, this was done in \cite{Ponsot:1999uf} using the theory of infinite dimensional representations of the quantum group $\mathcal{U}_q[\mathfrak{sl_2]}$. So, one would expect that the theory of infinite dimensional representations of more general quantum groups, especially those that lift to being representations of the quantum double (see \cite{frenkel2011positive} for some recent mathematical developments) would be relevant for the study of quantum Toda field theory. A closely related point of view would be the one from quantum Teichmuller theory for Liouville \cite{Teschner:2003em,Vartanov:2013ima} and generalizations thereof, namely that of higher Teichmuller theories \cite{hitchin1992lie,Biswas:1995ut,fock2006moduli}. The partition functions  analyzed here have also been described from the point of view of topological strings \cite{Vafa:2012fi}. Yet another connection to explore in detail would be that between the setup considered here and the geometric Langlands program with tame ramification \cite{beilinsonquantization,Kapustin:2006pk,frenkel2006local,Gukov:2006jk,frenkel2007langlands, Teschner:2010je,Tan:2013tq}. But, these are left for future considerations.
\section{Acknowledgements}
I would like to thank J. Distler for critical comments at various stages of this project. I also thank O. Chacaltana for many discussions on Gaiotto duality and the participants of the Brown Bag seminars for constructive feedback when this work was presented while it was still in progress. This material is based upon work supported by the National Science
Foundation under Grant Numbers PHY-1316033 and PHY-0969020. A part of this work was done during the String-Math 2013 conference at the Simons Center. I would like to thank its organizers for their hospitality and for putting together this stimulating conference. The Young diagrams in this paper were produced using the package ytableux.

\appendix

\section{Behaviour of functional determinants under scaling}
\label{appendixscaling}
Zeta function regularization is often used in the determination of functional determinants. The general strategy is the following. 
Let $A$ be the operator of interest. Forming a zeta function using the eigenvalues $A$ :
\begin{equation*}
\zeta^A(s) = \sum_n (\lambda_n)^{-s}.
\end{equation*}
This is typically convergent for $s > \sigma$ for some $\sigma \in \mathbb{R}$.  In many cases, this function can be analytically continued to arbitrary values of $s$ upto some poles that are away from $s=0$. This allows us to write the product of eigenvalues (formally) as
\begin{equation*}
\zeta^{A'}(0) = - \log (\prod_n \lambda_n) .
\label{loglambda}
\end{equation*}
Inverting this identity give us the regularized value for $\det(A)$
\begin{equation*}
\det(A)=\prod_n \lambda_n = e^{-\zeta^{A'}(0)}.
\end{equation*} 
Such regularizations often find use in problems that involve evaluating Gaussian path integrals on curved manifolds. In such cases, $A$ is typically an elliptic or a transversally elliptic operator that occurs in the quadratic part of the action. 

Let us now consider a scale transformation that changes the metric as $\tilde{g} = k^{-1}g$ and leads to a change in the eigenvalues as $\tilde{\lambda_n} = k \lambda_n$. The zeta function built out of $\tilde{\lambda_n}$ is related to the original one by
\begin{equation*}
\zeta^{A_k}(s) = k^{-s} \zeta^A(s).
\end{equation*}
Writing a regularized form of $\det(A_k)$ in terms of the original zeta function now requires an additional (=anomalous) term in the analogue of  (\ref{loglambda}),
\begin{equation*}
\zeta^{A'}(0) - (\log k) \zeta^A(0) = - \log (\prod_n \tilde{\lambda_n}).
\end{equation*}
Inverting this,
\begin{equation*}
\det(\tilde{A})=\prod_n \tilde{\lambda_n} = k^{\zeta^A(0)} e^{-\zeta^{A'}(0)}.
\end{equation*}
Factors of the form $k^{\zeta^A(0)}$ play an important role in the body of the paper.

\section{Special function redux}
\label{appendixspecial}

Some properties of the special functions that are used in the main body of the paper are collected here.  For a more detailed treatment of the analytical properties of these functions and a summary of the identities they obey, see \cite{spreafico2009barnes}.
The Barnes double zeta function and the Hurwitz zeta function have the following sum representations
\begin{eqnarray}
\zeta^B_2 (s;a,b,x)  &=& \sum_{m,n=0} (am + bn + x)^{-s},\\
\zeta^H (s,x) &=& \sum_{m=0} (n + x)^{-s}.
\end{eqnarray}
The derivatives at $s=0$ of these zeta functions are related to $\Gamma_2(x)$ and $\Gamma(x)$ in the following way, 
\begin{equation}
\zeta'^B_2(0;a,b,x) = \log (\Gamma_2 (x;a,b)) + \text{const},
\end{equation}
\begin{equation}
\zeta_H'(0,x) = \log (\Gamma(x)) + \text{const}.
\end{equation}
The $\Upsilon$ function that is often used in Liouville/Toda theory is defined as 
\begin{equation}
\Upsilon (x; b, b^{-1}) = \frac{1}{\Gamma_2 (x;b,b^{-1}) \Gamma_2 (Q-x;b,b^{-1})},
\end{equation}
where $Q = b + b^{-1}$. The derivative of the $\Upsilon$ function at $x=0$ also plays an important role in the DOZZ/FL correlators. It is given by, 
\begin{equation}
\Upsilon_0 = \frac{d \Upsilon(x)}{dx} \vert_{x=0} = \Upsilon (b),
\end{equation}
where the final equality follows from the asymptotic properties of $\Upsilon(x)$ \cite{Thorn:2002am,Fateev:2005gs}.
 Under a scaling transformation, $\Upsilon(x)$ has the following behaviour (this follows from the discussion in Section 2) ,
\begin{equation}
\Upsilon(\mu x ; \mu \epsilon_1 ,\mu \epsilon_2) = \mu^{2 \zeta_2^B(0,x;\epsilon_1,\epsilon_2)} \Upsilon(x ; \epsilon_1 ,\epsilon_2),
\end{equation}
with 
\begin{equation}
\zeta_2^B(0,x;\epsilon_1,\epsilon_2) = \frac{1}{4} + \frac{1}{12} \bigg(\frac{\epsilon_1}{\epsilon_2} \bigg) - \frac{x}{2} \bigg( \frac{1}{\epsilon_1} + \frac{1}{\epsilon_2} \bigg) + \frac{x^2}{2 \epsilon_1 \epsilon_2} .
\end{equation}
As a shorthand, let us summarize the above scaling behaviour by saying that the scale factor for $\Upsilon(x,\epsilon_1,\epsilon_2)$ (denoted by $\mu[\Upsilon(x,\epsilon_1,\epsilon_2)] $) is $2\zeta_2^B(0,x;\epsilon_1,\epsilon_2)$.
The Barnes G function (for $b=1$) can be related to the double gamma function defined above using (see Prop 8.5 in \cite{spreafico2009barnes} )
\begin{equation}
G(1+x) = \frac{\Gamma(x)}{\Gamma_2 (x;1,1)}
\end{equation}
Rewriting the above relationship in terms of derivatives of the Barnes double zeta and the Hurwitz zeta functions,
\begin{equation*}
e^{-\zeta_2^{B'} (0,x;1,1)+\zeta^{H'}(0,x;1,1)} = G (1+x).
\end{equation*}
Noting that,
\begin{equation}
\Upsilon(\frac{Q}{2} + i x) = \frac{1}{\Gamma_2(\frac{Q}{2}+ix) \Gamma_2(\frac{Q}{2}-ix)}
\end{equation}
The $H$ function and the $\Upsilon$ function are related to the Barnes $G$ function by
\begin{eqnarray}
H(x) &=& G(1+x) G(1-x), \\
\Upsilon_{b=1}(x) &=& \frac{G(1+x) G(3-x)}{\Gamma(x) \Gamma(2-x)} \\
\Upsilon_{b=1} (Q/2 + ix) &=& \frac{G(2+ix)G(2-ix)}{\Gamma(1+ix) \Gamma(1-ix)}
\end{eqnarray}
From Section 2, the scale factor for the $H$ function (specialized to $\epsilon_1= \epsilon_2=1$) is given by,
\begin{equation}
\mu [H(x)] = 2\zeta_2^B(0,x;1,1) - 2\zeta_H (0,x) =  - \frac{1}{6} + x^2 ,
\end{equation}
while the scale factor for the $\Upsilon$ function (again specialized to $\epsilon_1 = \epsilon_2 =1$) is
\begin{equation}
\mu [\Upsilon(x)] = 2\zeta_2^B(0,x;1,1) = \frac{5}{6} - 2x + x^2 = - \frac{1}{6} + (1-x)^2 .
\end{equation}

\section{Review of $\mathfrak{sl_2}$ embeddings}

Here, some important aspects of the theory of $\mathfrak{sl_2}$ embeddings in a complex lie algebra are collected. This is done here purely to introduce the notation and terminology. For the classical theory, consult \cite{malcev1950semi,dynkin1957semisimple,kostant1959principal,Lorente:1972xw} and for comprehensive textbook treatments, see \cite{collingwood1993nilpotent,carter1985finite}. The application of this theory in the context of reductions of WZW models has a long history. For a sample, see \cite{Feher:1992ed,deBoer:1993iz,Frappat:1992bs}.

By the theorem of Jacobson-Morozov, nilpotent orbits in complex semisimple lie algebras (upto conjugacy) are identified with inequivalent $\mathfrak{sl_2}$ embeddings (again, upto conjugacy). So, this allows the two objects to be used interchangeably and this freedom is used quite generously in the body of the paper. There are only a finite number of such inequivalent nilpotent orbits in any lie algebra. For the lie algebra $\mathfrak{sl}_N$, these orbits are indexed by partitions of N. For the other classical algebras, there is a still a partition type classification but one has to use appropriate notions of B,C or D- partitions. For the case of exceptional lie algebras, the list is known by explicit case-by-case constructions. 

\begin{figure} [!h]
\begin{center}
\begin{tikzpicture}
  \node (max) at (0,8) {$[6]$};
  \node (a) at (0,7) {$[5,1]$};
  \node (b) at (0,6) {$[4,2]$};
  \node (c) at (-1,5) {$[4,1^2]$};
  \node (d) at (1,5) {$[3^2]$};
  \node (e) at (0,4) {$[3,2,1]$};
  \node (f) at (-1,3) {$[3,1^3]$};
  \node (g) at (1,3) {$[2^3 ]$};
  \node (h) at (0,2) {$[2^2,1^2]$};
  \node (i) at (0,1) {$[2,1^4]$};
  \node (j) at (0,0){$[1^6]$};
  \draw[preaction={draw=white, -,line width=6pt}] (max) -- (a)--(b)--(c)--(e)--(f)--(h)--(i)--(j);
  \draw[preaction={draw=white, -,line width=6pt}] (b) -- (d)--(e);
  \draw[preaction={draw=white, -,line width=6pt}] (e) -- (g)--(h);
 \end{tikzpicture}
\end{center}
\caption{Hasse diagram for nilpotent orbits in $\mathfrak{sl}_6$.}
\label{hassesl6}
\end{figure}
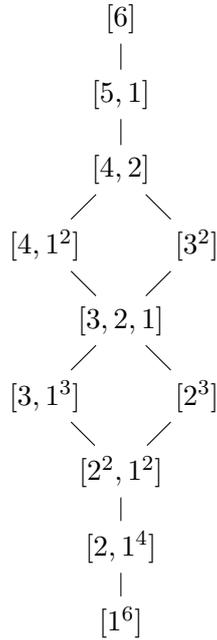

In all of these cases, a very useful way to identify a nilpotent orbit uniquely is by associating to it a weighted Dynkin diagram. This is done by relating nilpotent orbits to distinguished semi-simple orbits.  To every distinguished semi-simple orbit, one can attach a semi-simple element $h$ such that $(h,\alpha_i) \in {0,1,2}$ for $\alpha_i$ a simple root.  $h$ is called the Dynkin element and the diagram obtained by attaching the values  $(h,\alpha_i) $ to the corresponding nodes of the Dynkin diagram, the weighted Dynkin diagram. Note here that the number of nilpotent orbits is $\textit{much}$ less than $3^{\text{rank(G)}}$. So, not all possible assignments of the numbers ${0,1,2}$ are realized at a given node. There are standard ways to obtain the Dynkin element from the partition associated to a particular nilpotent orbit. The closure ordering on the nilpotent orbits defines a natural partial order on the set of nilpotent orbits. This can be translated to a partial order on the associated partition labels in the classical cases. For the case of $\mathfrak{sl}_N$, this is the usual dominance order on partitions. As examples of this partial order, the Hasse diagrams for nilpotent orbits in $\mathfrak{sl}_4$ and $\mathfrak{sl}_6$ are given in Figs \ref{hassesl4}, \ref{hassesl6}.

\begin{figure} [!h]
\begin{center}
\begin{tikzpicture}
  \node (max) at (0,4) {$[4]$};
  \node (a) at (0,3) {$[3,1]$};
  \node (b) at (0,2) {$[2^2]$};
  \node (c) at (0,1) {$[2,1^2]$};
  \node (d) at (0,0) {$[1^4]$};
  \draw[preaction={draw=white, -,line width=6pt}] (max) -- (a) -- (b)--(c) --(d) ;
\end{tikzpicture}
\end{center}
\caption{Hasse diagram for nilpotent orbits in $\mathfrak{sl}_4$.}
\label{hassesl4}
\end{figure}
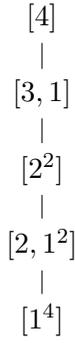

Finally, the complex dimension of a nilpotent orbit in type $A$ has a simple formula. Let $P$ be a partition of $n$. Consider the nilpotent orbit of partition type $P$ in the lie algebra $\mathfrak{g}=A_{n-1}$. Let, $P^t = [r_1, r_2 \ldots]$ be the transpose partition. Then,
\begin{equation}
\text{dim}_{\mathbb{C}}(\mathcal{O}_{P}) = n^2 - \sum r_i^2.
\end{equation}
\section{Conformal Bootstrap}
\label{bootstrap}
It is useful to recall how the conformal bootstrap procedure proceeds for Liouville theory. The basic idea is the procedure put forward in BPZ (for a detailed review, see \cite{Zamolodchikov:1990ww} ). For a modern understanding of the analytical bootstrap procedure as it is applies to the case of Liouville CFT, see \cite{Teschner:2001rv}.

Let us start with the two point function on the sphere. Conformal invariance constrains this to be of the form
\begin{equation*}
V_{0,2} = \langle \mathcal{O}_\alpha  \mathcal{O}_\beta  \rangle = \frac{\delta_{\alpha \beta}}{|z_1-z_2|^{\Delta}}
\end{equation*}

The three point function is similarly constrained but not completely determined by requirements of conformal invariance. 
\begin{equation*}
V_{0,3} = C(\alpha_1,\alpha_2,\alpha_3) |z_{12}|^{-2(\Delta_1 + \Delta_2 -\Delta_3)} |z_{13}|^{-2(\Delta_1 + \Delta_3 - \Delta_2)} |z_{23}|^{-2(\Delta_2 + \Delta_3 -\Delta_1)}
\end{equation*}

The dynamics of the theory is encoded in $C(\alpha_1,\alpha_2,\alpha_3)$. The procedure of conformal bootstrap outlined in BPZ, \cite{Zamolodchikov:1990ww} starts with the writing of the general four point function in terms of the three point functions and a special function known as the conformal block. 

Let us start with a generic four point function and insert a complete set of states in between the four operators.
\begin{equation}
\langle \mathcal{O}_{\alpha_1} \mathcal{O}_{\alpha_2} \mathcal{O}_{\alpha_3} \mathcal{O}_{\alpha_4} \rangle = \sum_{[\alpha]} \text{or} \int_{[\alpha]} \langle \mathcal{O}_{\alpha_1} \mathcal{O}_{\alpha_2} \mathcal{O}_{[\alpha]} \rangle \langle {\mathcal{O}_{[\alpha]}}^* \mathcal{O}_{\alpha_1} \mathcal{O}_{\alpha_2}  \rangle 
\label{insertion}
\end{equation}
where $[\alpha]$ denotes the conformal family associated to a primary $\mathcal{O}_\alpha$. Note that the members of the conformal family can be obtained by acting with the operators $\mathcal{L}_{-m}$ ($m > 0$). Both symbols $\sum$ or $\int$ are included to highlight the the fact that in arbitrary cases, there may be a continuous integral and a discrete sum involved. However, it is the integral sign that is employed in most parts of the paper. This is done to simplify notation.

Now, one can proceed by using the OPE between the first two operators to write the first term in the following way
\begin{equation*}
\mathcal{O}_{\alpha_1} \mathcal{O}_{\alpha_2} = \int d \alpha C(\alpha_1, \alpha_2, \alpha) z^{\Delta_\alpha - \Delta_{\alpha_1}- \Delta_{\alpha_2}}\bar{z}^{\bar{\Delta}_\alpha - \bar{\Delta}_{\alpha_1}- \bar{\Delta}_{\alpha_2}} \mathcal{O_{[\alpha]}}
\end{equation*} 
where,
\begin{equation*}
\mathcal{O}_{[\alpha]} = \mathcal{O}_\alpha  + \Omega_{12}^{\alpha,1} z L _{-1} \mathcal{O}_\alpha + \bar{\Omega}_{12}^{\alpha,1} \bar{z} \bar{L} _{-1} \mathcal{O}_\alpha + \Omega_{12}^{\alpha,\{1,1\}} z^2 L _{-1}^2 \mathcal{O}_\alpha + \ldots.
\end{equation*}
The dynamics of the theory is encoded in the coefficients $\Omega_{12}^{\alpha,\{ \ldots \}}$ and $\bar{\Omega}_{12}^{\alpha,\{ \ldots \}}$ that appear in the above expansion. These constants obey a recursive set of linear equations which can be solved level by level. The final solution for $\Omega_{12}^{\alpha,\{ \ldots \}}$ at some low levels have the following form
\begin{eqnarray*}
\Omega_{12}^{\alpha,\{1 \}} &=& \frac{\Delta_{\alpha} - \Delta_{\alpha_1} - \Delta_{\alpha_2} }{2 \Delta_{\alpha} },\\
\Omega_{12}^{\alpha,\{1,1\}} &=& \frac{(\Delta_{\alpha} - \Delta_{\alpha_1} - \Delta_{\alpha_2})(\Delta_{\alpha} - \Delta_{\alpha_1} - \Delta_{\alpha_2}+1)}{4 \Delta_{\alpha} (2 \Delta_{\alpha} +1 )} - \frac{3}{2(\Delta_{\alpha}+1)} \Omega_{12}^{\alpha,\{1 \}}.
\end{eqnarray*} 

As a simple example, consider the three point function in Liouville CFT.

\subsection{$V_{(0,3)}=V[\mathfrak{sl_2}]_{0,([1^2],[1^2],[1^2])}$}
In the AGT correspondence, this is the correlator assigned to a theory of four free hypermultiplets. By DOZZ, we have
\begin{equation*}
 V[\mathfrak{sl_2}]_{0,([1^2],[1^2],[1^2])} = C(\alpha_1,\alpha_2,\alpha_3) |z_{12}|^{-2(\Delta_1 + \Delta_2 -\Delta_3)} |z_{13}|^{-2(\Delta_1 + \Delta_3 - \Delta_2)} |z_{23}|^{-2(\Delta_2 + \Delta_3 -\Delta_1)},
\end{equation*}
where $C(\alpha_1,\alpha_2,\alpha_3)$ is given by
\begin{eqnarray*}
 C(\alpha_1,\alpha_2,\alpha_3) &=&  \bigg[ \pi \mu \gamma(b^2) b^{2-2b^2} \bigg]^{(Q-\sum_i \alpha_i)/b} \times \\ &&\frac{\Upsilon(b)\Upsilon(2 \alpha_1)\Upsilon(2\alpha_2)\Upsilon(2\alpha_3)}{\Upsilon(\alpha_1 + \alpha_2 + \alpha_3 - Q)\Upsilon(\alpha_1 + \alpha_2 - \alpha_3)\Upsilon(\alpha_2 + \alpha_3 - \alpha_1)\Upsilon(\alpha_3 + \alpha_1 - \alpha_2)}
\end{eqnarray*}
%

Note that $\Upsilon(x)$ is an entire function except for zeros at $x= -m b -n b^{-1}$ or $x=Q+m'b + n'b^{-1}$ for $m,n,m',n' \in \mathbb{Z} ^{\ge0}$. The DOZZ three point function then has a pole when any one of the following conditions is satisfied,
\begin{eqnarray*}
 \alpha_1 + \alpha_2 + \alpha_3 - Q = \Omega_{m,n}, \\
\alpha_1 + \alpha_2 - \alpha_3 = \Omega_{m,n}, \\
\alpha_2 + \alpha_3 - \alpha_1  = \Omega_{m,n}, \\
\alpha_3 + \alpha_1 - \alpha_2 = \Omega_{m,n},
\end{eqnarray*}
where $\Omega_{m,n}$ is used to denote the string of points $-m b - n b^{-1}$ and $Q+m'b+n'b^{-1}$. The set of poles matches with the screening conditions that arise from doing the path integral of the Liouville zero modes. Let us recall the general form of a screening condition for future purposes. 
\begin{equation*}
 \sum_i \alpha_i + (g-1) Q = \Omega_{m,n},
\end{equation*}
where $g$ is the genus and the sum is over all punctures. 
Starting with any one of the conditions, the other three can be obtained by single Weyl reflections $W_i : \alpha_i \rightarrow Q - \alpha_i$. Observe that overall Weyl reflections do not give a new screening condition. For example, starting with the condition $\sum \alpha -Q = \Omega_{m,n}$ and reflecting using $W: \sum \alpha \rightarrow Q - \sum \alpha$ leads to the same screening condition. This implies that the total number of screening conditions is four and not eight.
 Now, using the AGT primary map, the screening conditions can be rewritten in terms of the mass deformations
\begin{eqnarray}
 \frac{Q}{2} + m_1 + m_2 + m_3 = \Omega_{m,n}, \\
\frac{Q}{2} + m_1 + m_2 - m_3 = \Omega_{m,n}, \\
\frac{Q}{2} +  m_2 + m_3 - m_1  = \Omega_{m,n}, \\
\frac{Q}{2} + m_3 + m_1 - m_2= \Omega_{m,n}.
\end{eqnarray}
Observe that when any one of the hypermultiplet masses is set to zero, there is no pole since the point $Q/2$ does not belong to the string of poles $\Omega_{m,n}$ unless $Q=0$. $Q=0$ is possible only if $b=\pm i$. One can not naively continue the result to pure imaginary values of $b$ since that is outside the region of analyticity of the DOZZ three point function \cite{Zamolodchikov:2005fy,Harlow:2011ny}. Since flat directions in the moduli space are opening up when such relations are satisfied, one would naively expect $Z_{\mathbb{S}^4}$ to diverge. But, such a direct interpretation for the pattern of divergences does not seem to be possible. The mass relations are instead encoded in the polar divisors of the integrand for  $Z_{\mathbb{S}^4}$ in a $Q$-deformed manner. It is not immediately clear as to what physical meaning should be attributed to the lattice of poles. But, there is still something useful that one can learn from this simple example of a three point function. Namely, the number of hypermultiplets is nothing but the total number of 
screening conditions . This simple relation between number of screening conditions and $n_h$ holds for all the free theories.  
%
The bootstrap program entails using insertions of complete states as in (\ref{insertion}) and obtaining all higher point functions starting from the three point function. Requiring that the resulting higher point functions (on arbitrary genus surfaces) obey the crossing relations and its generalizations ends up being a very strong constraint on the three point function that it determines its analytical structure. One can work in the opposite direction as well. This would imply starting with the DOZZ three point function and then checking that the higher point functions have the required pole structure and obey crossing relations. In the example below, we will see how bootstrap produces the required pole structure as the result of an intricate interplay of various different factors. One could, ultimately, hope to understand Toda bootstrap at this level of detail.
\subsection{$V_{(0,4)}=V[\mathfrak{sl_2}]_{0,([1^2],[1^2],[1^2],[1^2])}$}
This is the correlator corresponding to $\mathcal{N} =2$ SYM with gauge group $SU(2)$ and $N_f=4$. The flavor symmetry for this theory is $SO(8)$. The theory has four mass deformation parameters which can each be assigned to a $SU(2)$ flavor subgroup of $SO(8)$. These mass parameters will be related to the Liouville momenta in the following fashion
\begin{equation*}
 \alpha_i = \frac{Q}{2} + m_i
\end{equation*}
The eigenvalues of the mass matrix are $m_1 + m_2$, $m_1 - m_2$, $m_3+ m_4$ and $m_3 - m_4$. 
To write down the four point function in Liouville theory, one usually takes  $\alpha_i, \alpha$ to lie on the physical line. That is, $\alpha_i = Q/2 + i s_i^+, \alpha =Q/2 + is^+$ for $s_i^+, s^+ \in \mathbb{R}^{+}$. The four point function can then be written as
\begin{eqnarray*}
 Z_{S^4} = && V_{0,4}(\alpha_1,\alpha_2,\alpha_3,\alpha_4) = \\ &&\int_{\alpha \in {\frac{Q}{2}+ i s^+}}d \alpha C(\alpha_1, \alpha_2, \alpha)  C (Q - \alpha,\alpha_3,\alpha_4) \mathcal{F}_{12}^{34} (c, \Delta_{\alpha}, z_i) \mathcal{F}_{12}^{34}(c, \Delta_{Q -\alpha}, \bar{z}_i)
\label{fourpoint1}
\end{eqnarray*}
The fact that $\alpha \in {\frac{Q}{2}+ i s}$ implies $\bar{\alpha}=Q-\alpha$ has been used in the above equation. Now, using the symmetry of the entire integrand under the Weyl reflection $\alpha \rightarrow Q -\alpha $,  the integral can be unfolded to one over $\mathbb{R}$. This gives 
\begin{equation*}
 V_{0,4}(\alpha_1,\alpha_2,\alpha_3,\alpha_4) = \frac{1}{2} \int_{\alpha \in {\frac{Q}{2}+ i s}}d \alpha C(\alpha_1, \alpha_2, \alpha)  C (Q - \alpha,\alpha_3,\alpha_4) \mathcal{F}_{12}^{34} (c, \Delta_{\alpha}, z_i) \mathcal{F}_{12}^{34}(c, \Delta_{Q -\alpha}, \bar{z}_i)
\end{equation*}
where $s \in \mathbb{R}$.
Now, observe that the integrand depends just on $\alpha$ and not on $\bar{\alpha}$. This allows us to analytically continue the integrand to arbitrary values of $\alpha$ and then interpret (\ref{fourpoint1}) as a contour integral. Let us now study the analytical structure of the four point function by looking at different parts of the integrand (see \cite{Teschner:2001rv}).
\begin{enumerate}
 \item Although the \textit{Vir} conformal blocks are completely constrained by symmetry, no closed form expression is known. But,  its analytical properties wrt $\alpha$ are deduced by observing that the conformal blocks can be written as
\begin{equation*}
  \mathcal{F} (c, \Delta_i, \Delta_\alpha, z_i) = z_{13}^{-2(\Delta_1 + \Delta_2 + \Delta_3 - \Delta_4)} z_{14}^{-2(\Delta_1 + \Delta_4 -\Delta_2 -\Delta_3)} z_{24} ^{-4\Delta_2} z_{34}^{-2(\Delta_3 + \Delta_4 - \Delta_1 - \Delta_2)} F (c, \Delta_i , \Delta_\alpha, q)
\end{equation*}
where $q = z_{12} z_{34} / z_{13} z_{24}$. $F (c, \Delta_i , \Delta_\alpha, q)$ has the following series expansion 
\begin{equation*}
F (c, \Delta_i, \Delta_\alpha, q) = q^{\Delta_\alpha - \Delta_1 - \Delta_2} \sum_{i=0}^{\infty} F_i(c, \Delta_\alpha, \Delta_i) q^i 
\end{equation*}
Each term in the expansion can in turn be written as a ratio of two polynomials. 
\begin{equation*}
 F_i = \frac{P_i (c,\Delta , \Delta_i)}{Q_i(c,\Delta)}
\end{equation*}
The denominator $Q(c, \Delta_\alpha)$ is nothing but the divisor of the Kac determinant at level $i$.  It is zero when when $\alpha$ takes values corresponding to degenerate representations
\begin{equation*}
 \alpha = -  \frac{(m+1) b}{2} -  \frac{(n+1) b^{-1}}{2}.
\end{equation*}
When this condition is satisfied, there is a null vector in the Verma module at level $(m+1)(n+1)$. The zero of $Q(x,\Delta_\alpha)$ leads to a pole for $\mathcal{F}(z)$. A similar sequence of arguments show that at exactly the same values of $\alpha$, $\mathcal{F}(\bar{z})$ also picks up a pole. This is because $\Delta_\alpha $ = $\Delta_{Q-\alpha}$ and the dependence of the chiral and the anti-chiral conformal blocks on $\alpha$ is only through their dependence on $\Delta_\alpha$.
 So, $\mathcal{F}(z)$ and $\mathcal{F}(\bar{z})$ combine to give a double pole. However, the factor $\Upsilon(2 \alpha) \Upsilon ( 2(Q-\alpha))$ has a double zero exactly at these values. So, they cancel.
\item The $\Upsilon$ functions in the denominator (from both two $C(\ldots)$ factors combined) have simple poles when any one of the  following conditions are satisfied
\begin{eqnarray*}
 \alpha_1 + \alpha_2 + \alpha &=& Q -\Omega_{m,n} \hspace{1in} \alpha_1 + \alpha_2 + \alpha = 2Q+\Omega_{m,n} \\
 \alpha_1 + \alpha_2 - \alpha &=& -\Omega_{m,n} \hspace{1in} \alpha_1 + \alpha_2 - \alpha = Q+\Omega_{m,n} \\
 \alpha_1 + \alpha - \alpha_2 &=& -\Omega_{m,n} \hspace{1in} \alpha_1 + \alpha - \alpha_2 = Q+\Omega_{m,n} \\
 \alpha_2 + \alpha - \alpha_1 &=& -\Omega_{m,n} \hspace{1in}  \alpha_2 + \alpha - \alpha_1  = Q+\Omega_{m,n} \\
 \alpha_3 + \alpha_4 - \alpha &=&  -\Omega_{m,n} \hspace{1in} \alpha_3 + \alpha_4 - \alpha  = Q + \Omega_{m,n} \\
 \alpha_3 + \alpha_4 + \alpha &=& Q -\Omega_{m,n} \hspace{1in}  \alpha_3 + \alpha_4 + \alpha  = 2 Q + \Omega_{m,n} \\
 \alpha_3 - \alpha - \alpha_4 &=& -Q -\Omega_{m,n} \hspace{1in}  \alpha_3 - \alpha - \alpha_4 = \Omega_{m,n} \\
 \alpha_4 - \alpha - \alpha_3 &=& -Q -\Omega_{m,n} \hspace{1in} \alpha_4 - \alpha - \alpha_3 = \Omega_{m,n} \\
\end{eqnarray*}
Let us fix $\Re(\alpha_i) = Q/2$. As we will momentarily see, the integral is well defined for arbitrary values of $\Im(\alpha_i)$. One can also continue to arbitrary values of $\Re(\alpha_i)$ except when they end up satisfying a screening condition. In those cases, poles emerge because the contour gets pinched. To see these aspects, it is better to change variables. Set $\alpha_i = Q/2 + i s_i$ where  $s_i \in \mathbb{R}$ The above set of equations then imply strings of poles at the following values in the $\alpha$-plane.
 \begin{eqnarray*}
  \alpha &=&  -\Omega_{m,n} -i(s_1 +s_2) \hspace{1in}   \alpha = Q+\Omega_{m,n}  -i(s_1 +s_2) \\
   \alpha &=& Q +\Omega_{m,n} + i(s_1 +s_2) \hspace{1in}   \alpha =  - \Omega_{m,n} + i (s_1 + s_2)\\
  \alpha  &=& -\Omega_{m,n} + i(s_2 -s_1) \hspace{1in} \alpha  = Q+\Omega_{m,n} + i(s_2 -s_1) \\
  \alpha &=& -\Omega_{m,n} + i (s_1 -s_2) \hspace{1in}   \alpha   = Q+\Omega_{m,n} - i(s_1 -s_2) \\
  \alpha &=& Q +\Omega_{m,n} + i(s_3 + s_4) \hspace{1in}   \alpha  =  - \Omega_{m,n} + i(s_3 + s_4) \\
  \alpha &=&  -\Omega_{m,n} - i(s_3 + s_4) \hspace{1in}    \alpha  =  Q + \Omega_{m,n} - i(s_3 + s_4) \\
  \alpha  &=& Q +\Omega_{m,n} + i(s_3 - s_4) \hspace{1in}    \alpha  = - \Omega_{m,n} + i(s_3 -s_4)\\
   \alpha  &=& Q +\Omega_{m,n} + i(s_4 - s_3) \hspace{1in}  \alpha  = - \Omega_{m,n} + i(s_4-s_3) \\
\end{eqnarray*}
Notice that every $\Upsilon$ function leads one string of left-poles (poles strictly in the region to the left of the contour) and another string of right-poles (pole strictly in the region right of the contour). It is useful to plot the poles in the $\alpha$ plane (See Fig \ref{liouville4pointfigure}). The blue line indicates the position of the contour while the green lines indicate that of the poles. Note that for irrational $b$, all poles occur at distinct points along the line. The green lines are drawn as continuous lines just for convenience. The point on the green lines that is closest to the contour is the location of the first pole.
\begin{figure}[h!]
\begin{center}
\includegraphics[scale=1.5]{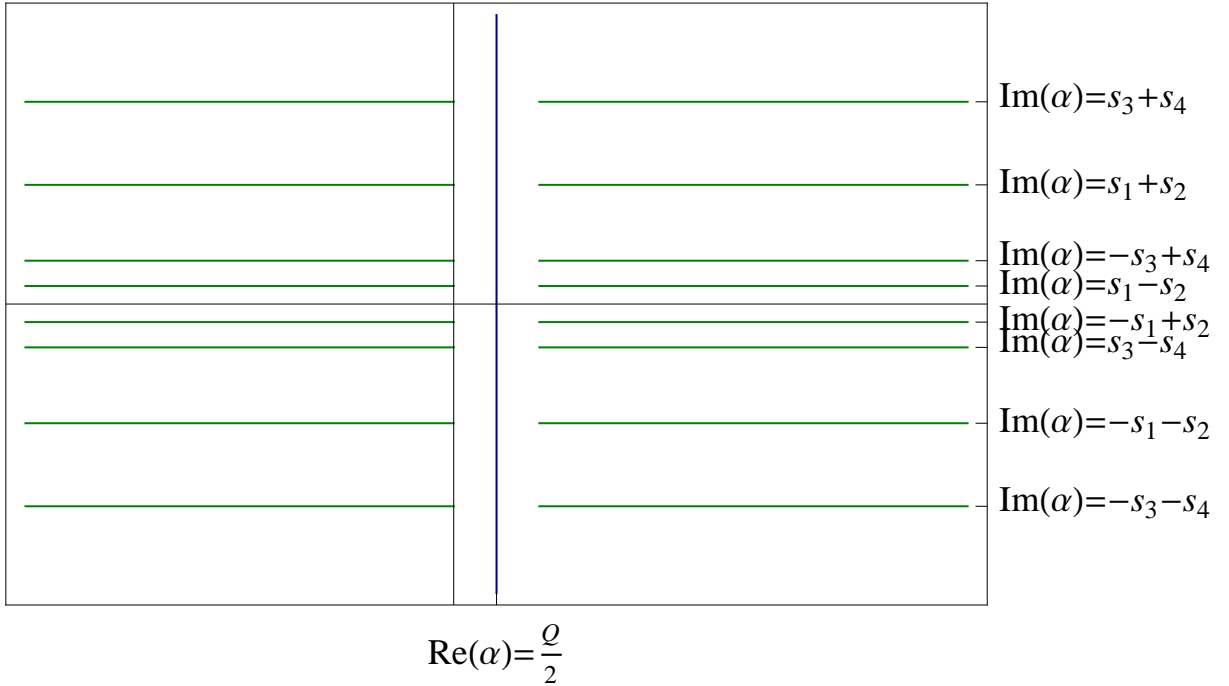}
\end{center}
\caption{Analytical structure of the integrand for $V_{0,4}$}
\label{liouville4pointfigure}
\end{figure}

It is useful to define an object called the set of all polar divisors of the integrand,
\begin{equation*}
 \mathcal{D}_i \equiv \{ \Im(\alpha) = k | k \in \{s_1 + s_2, -s_1 - s_2, s_1 - s_2, s_2 - s_1, s_3 + s_4, -s_3 -s_4, s_3 - s_4, s_4-s_3\} \}. 
\end{equation*}

To define the continuation to arbitrary values of $\alpha_i$, it is important to note that the poles are away from the contour as long as the following conditions are satisfied,
\begin{eqnarray}
|\Re(\alpha_1 - \alpha_2)| & <& Q /2 ,\\
|\Re(Q - \alpha_1 - \alpha_2)| &<& Q /2 ,\\
|\Re(\alpha_3 - \alpha_4)| &<& Q /2 ,\\
|\Re(Q- \alpha_3 - \alpha_4)| &<& Q /2.
\end{eqnarray}
When going outside the range allowed by these inequalities, one should watch for poles to cross the contour and indent the contour correspondingly.  This new contour can be rewritten as the original contour plus a finite number of circles around the poles that crossed. There are a finite number of extra terms corresponding to the residues at these poles. This prescription suffices as long as all the polar divisors $\mathcal{D}_i$ are distinct. When some of them align, the contour can get pinched when $\alpha_i$ takes arbitrary values. Let us called the divisors that align as $\mathcal{D}_1 \& \mathcal{D}_2$. The pinching happens when the left poles in $\mathcal{D}_1$ have moved a distance  $\ge Q/2$ to the right while simultaneously, the right poles of $\mathcal{D}_2$ have moved by a distance $\ge Q/2$ to the left. If there are no new zeros emerging, such pinching leads to poles in the integral. In some cases, new zeros do emerge. The poles that arise when conditions of the form $s_i + s_i = s_i - s_j$, where $(i,j)$ is either $(1,2)$ or $(2,3)$, are satisfied are canceled by the zeros of $\Upsilon(2 \alpha_1), \Upsilon(2 \alpha_2), \Upsilon(2 \alpha_3), \Upsilon(2 \alpha_4)$. But, others (say, those that follow from $s_1 + s_2 = s_3 + s_4$) will remain as poles of the integral. These are exactly the cases for which the screening condition is satisfied. As expected, the four point function has simple poles only at these values. 

%
%
 \end{enumerate}

\bibliographystyle{utphys.bst}
\bibliography{scale.bbl}

\end{document}